\documentclass[11pt]{article}
\usepackage{geometry}                
\usepackage[parfill]{parskip}    
\usepackage{graphicx}
\usepackage{amssymb}
\usepackage{amsmath}
\usepackage{mathtools}
\usepackage{epsfig}
\DeclareMathOperator\erf{erf}
\usepackage{multirow}
\usepackage[square, comma, sort&compress, numbers]{natbib}

\title{Modularity Enhances the Rate of Evolution in a Rugged Fitness Landscape}
\author{Jeong-Man Park$^{1,2}$, Man Chen$^1$, Dong Wang$^1$,
and Michael W. Deem$^{1,3}$\\
\hbox{}$^1$Department of Physics \& Astronomy\\
Rice University, Houston, TX 77005--1892, USA\\
\hbox{}$^2$Department of Physics, The Catholic University of Korea, Bucheon
420-743, Korea\\
\hbox{}$^3$Center for Theoretical Biological Physics,
Rice University, Houston, TX 77005--1892, USA\\
}

\begin{document}
\maketitle

\begin{abstract}
Biological systems are modular, and this modularity affects the evolution
of biological systems over time and in different environments. We here 
develop a theory for the dynamics of evolution in a rugged, modular
fitness landscape.  We show analytically how horizontal gene
transfer couples to the modularity in the system and leads to
more rapid rates of evolution at short times.
The model, in general, analytically demonstrates a
selective pressure for the prevalence of modularity in biology.
We use this model to show
how the evolution of the influenza virus is affected by
the modularity of the proteins that are recognized
by the human immune system.  
Approximately 25\% of the observed rate of fitness increase of the virus
could be ascribed to a modular viral landscape.
\end{abstract}


\maketitle

\section{Introduction}

Biological systems are modular, and the organization of their
genetic material reflects this 
modularity \cite{Waddington,Simon,Hartwell1999,Rojas}.  
Complementary to this modularity is a set of evolutionary
dynamics that evolves the genetic material of biological systems.
In particular, 
horizontal gene transfer (HGT) is an important mechanism of
evolution, in which genes, pieces of
genes, or multiple genes are transferred from one individual to
another  \cite{hgt1,hgt,Goldenfeld2011a}.
Additionally,
multi-body contributions to the fitness function in biology 
are increasingly thought to be an important factor in evolution
\cite{Breen2012},
leading
to a rugged fitness landscape and glassy evolutionary dynamics.
The combination of modularity and horizontal gene transfer
provide an effective mechanism for evolution upon a rugged
fitness landscape \cite{Sun}.
The organization of biology into modules simultaneously
restricts the possibilities for function, because the modular
organization is a subset of all possible organizations, and may
lead to more rapid evolution, because the evolution
occurs in a vastly restricted modular subspace of all possibilities
\cite{Sun,Alon}.  Our results explicitly demonstrate
this trade off, with $t^*$ serving as the crossover time from the latter
to the former regime.

Thus, the fitness function in biology is increasingly realized to be rugged,
yet modular.  Nonetheless, nearly all analytical theoretical treatments assume
a smooth fitness landscape with a dependence only on Hamming distance 
from a most-fit sequence \cite{Park06}, a linear or multiplicative
fitness landscape for
dynamical analysis of horizontal gene transfer \cite{Levine,Shraiman},
or an uncorrelated random energy model \cite{Derrida1991,Peliti2011}.
Horizontal gene transfer processes on more general, but still smooth,
landscapes have been analyzed \cite{Park,Munoz2,DS1,DS2,DS3}.
Here we provide, to our knowledge, the first analytical
treatment of a finite-population Markov model of evolution
showing how horizontal gene transfer couples to modularity 
in the fitness landscape.  We prove analytically that
modularity can enhance the rate of evolution for rugged fitness landscapes
in the presence of horizontal gene transfer.  This foundational result
in the physics of biological evolution offers a clue to why biology
is so modular.  We demonstrate this theory with an application to
evolution of the influenza virus.

We introduce and solve a model of individuals evolving
on a modular, rugged fitness landscape.
The model is constructed to represent several fundamental
aspects of biological evolution: a finite population, mutation and
horizontal gene transfer, and a rugged fitness landscape.
For this model, we will show that the evolved fitness
is greater for a modular landscape than for a
non-modular landscape.  This result holds for $t<t^*$ where
$t^*$ is a crossover time, larger than typical
biological timescales.  The dependence of the evolved
fitness on modularity is multiplicative with the horizontal
gene transfer rate, and the advantage of
modularity disappears when horizontal gene transfer is
not allowed.  Our results describe the response of the system
to environmental change.  In particular, we show that
modularity allows the system
to recover more rapidly from change, and fitness values attained
during the evolved response to change increase with modularity
for large or rapid environmental change.

\section{Theory of the Rate of Evolution in a Rugged Fitness Landscape}

We use a Markov model to describe the evolutionary process.  There
are $N$ individuals.  Each individual $\alpha$ replicates at a rate
$f_\alpha$. 
The average fitness in the population is defined as
$\langle f \rangle = \frac{1}{N} \sum_{\alpha=1}^N f_\alpha$.
 Each individual has a sequence $S^\alpha$ that is composed of $L$
loci, $s_i^\alpha$.  For simplicity, we take $s_i^\alpha = \pm 1$.  Each of
the loci can mutate to the opposite state with rate $\mu$.
Each sequence is composed of $K$ modules of length $l = L/K$.
A horizontal gene transfer process randomly replaces
the $k$th module in the sequence of individual $\alpha$,
 e.g.\ the sequence loci $s_{(k-1) l + 1}^\alpha \ldots s_{k l}^\alpha$,
with the corresponding sequences
from a randomly chosen individual $\beta$ at rate $\nu$.  The
\emph{a priori} rate of sequence change in a population  is, therefore,
$N \mu L$ + $N \nu L / 2$.
Since the fitness landscape in biology is rugged, we use
a spin glass to represent the fitness:
\begin{eqnarray}
f[S] &=& 2 L + H[S]
\nonumber \\
H[S] &=& \sum_{ij} s_i s_j J_{ij} \Delta_{ij}
\label{1}
\end{eqnarray}
$J_{ij}$ is a quenched, Gaussian random matrix, with variance $1/C$.
As discussed in Appendix A, 
the offset value $2 L$ is chosen by Wigner's semicircle law so that the
minimum eigenvalue of $f$ is non-negative. 
The entries in the matrix $\Delta$ are zero or one, with probability
$C/L$ per entry, so that the average number of connections per row is $C$.
We introduce modularity by an excess of interactions in $\Delta$ along the
$l \times l$ block diagonals of the $L \times L$
connection matrix. There are $K$ of these
block diagonals.  Thus,
the probability of a connection is
$C_0/L$ when
$ \lfloor i/l \rfloor \ne \lfloor j/l \rfloor$
and $C_1/L$ when
$ \lfloor i/l \rfloor = \lfloor  j/l \rfloor$.  The number of
connections is $C = C_0   + (C_1 - C_0) /K$.
Modularity is defined by $M = (C_1 - C_0)  / (K C)$ and
obeys $-1/(K-1) \le M \le 1$.

The Markov process describing these evolutionary dynamics 
includes terms for replication ($f$), mutation ($\mu$), and
horizontal gene transfer ($\nu$):
\begin{eqnarray}
\frac{d P ( \{ n_{\bf a} \}; t)}{d t} &=&
\sum_{ \{ {\bf a} \} }
\bigg[
f(S_{\bf a}) (n_{\bf a}-1) \sum_{ \{ {\bf b} \ne {\bf a} \} }
 \frac{n_{\bf b}+1}{N} P(n_{\bf a}-1, n_{\bf b}+1; t)
- 
f(S_{\bf a}) n_{\bf a} \sum_{ \{ {\bf b} \ne {\bf a} \} } 
\frac{n_{\bf b}}{N} P(n_{\bf a}, n_{\bf b}; t)
\bigg]
\nonumber \\ &&
+ \mu
\sum_{ \{ {\bf a} \} }
\sum_{ \{ {\bf b}=\partial {\bf a} \} }
\bigg[
(n_{\bf b}+1)  P(n_{\bf a}-1, n_{\bf b}+1; t)
-
n_{\bf b}  P(n_{\bf a}, n_{\bf b}; t)
\bigg]
\nonumber \\ &&
+ \nu
\sum_{ \{ {\bf a} \} }
\sum_{k=1}^K
\sum_{ \{ {\bf b}, {\bf b}_k \ne {\bf a}_k  \} }
\bigg[
(n_{ {\bf a} / {\bf b}_k } +1)
 \frac{ n_ { {\bf b} / {\bf a}_k }  }{N} P(n_{\bf a}-1, 
n_{ {\bf a} / {\bf b}_k } +1; t)
\nonumber \\ &&
-
n_{ {\bf a} / {\bf b}_k }
 \frac{ n_ { {\bf b} / {\bf a}_k }  }{N} P(n_{\bf a}, 
n_{ {\bf a} / {\bf b}_k } ; t)
\bigg]
\nonumber \\ 
\label{2}
\end{eqnarray}
Here $n_{\bf a}$ is the number of individuals with sequence $S_{\bf a}$, with
the vector index ${\bf a}$ used to label the $2^L$ sequences.
This process conserves $N = \sum_{\bf a} n_{\bf a}$.
The notation $\partial {\bf a}$ means the $L$ sequences created by
a single mutation from sequence $S_{\bf a}$.
The notation ${\bf a} / {\bf b}_k$ means the sequence created by
horizontally gene transferring module $k$ from sequence 
$S_{\bf b}$ into sequence $S_{\bf a}$.

We consider how a population of initially random sequences
adapts to a given environment, averaged over the distribution
of potential environments.
For example, in the context of influenza evolution,
these sequences arise, essentially randomly, by transmission
from swine.
As discussed in Appendix B,
a short-time expansion for the average fitness can be derived
by recursive application of this master equation:
\begin{eqnarray}
\langle f(t) \rangle &=& 2 L + a t + b t^2
\nonumber \\ 
a &=&  2 L \left( 1 - \frac{1}{N} \right)
\nonumber \\ 
b &=& 
- \frac{4 L^2}{N} \left( 1-\frac{1}{N}\right)
- 4 \mu L \left( 1-\frac{1}{N}\right)
\nonumber \\ && 
- 2 \nu L \left[
\left( 1-\frac{1}{K}\right)
\left( 1-M \right)
\left( 1-\frac{4}{N}\right)
+
\frac{1}{N}
\right]
\left( 1-\frac{1}{N}\right) 
\label{3}
\end{eqnarray}
Result (\ref{3}) is exact for all finite $N$.
Note that the effect of modularity enters at the quadratic level and 
requires a non-zero rate of horizontal gene transfer, $\nu > 0$.
We see that $\langle f_{M>0}(t) \rangle > \langle f_0(t) \rangle$ 
for short times.

From the master equation, we also calculate the  sequence
divergence, defined as $D = \frac{1}{N}\sum_{\alpha=1}^N
\langle
\frac{L - S_\alpha(t) \cdot S_\alpha(0)}{2}
\rangle
$.
As discussed in Appendix C, recursive application of Eq.\ (\ref{2})  gives
\begin{eqnarray}
D &=& \alpha t + \beta t^2 
\nonumber \\ 
\alpha &=& 
L^2 \left( 1 - \frac{1}{N} \right)
+ \mu L + \frac{\nu L}{2} \left( 1 - \frac{1}{N} \right)
\nonumber \\ 
\beta &=&
- L^3 
 \left( 1 - \frac{1}{2 L} + \frac{3}{2 L N} \right)
\left( 1 - \frac{1}{N} \right)
-2 \mu L^2 \left( 1 - \frac{1}{N} \right)
- \nu L^2 \left( 1 - \frac{1}{N} \right)
\nonumber \\ &&
- \mu^2 L 
-\mu \nu L \left( 1 - \frac{1}{N} \right)
- \frac{1}{4} \nu^2 L \left( 1 - \frac{1}{N} \right)
\label{4}
\end{eqnarray}
The sequence divergence does not depend on modularity at second order in 
time.  Note the terms at order $t^n$ not proportional $\mu^m \nu^{n-m}$
are due to discontinuous changes in sequence resulting from the
fixed population size constraint.

Introduction of a new sequence into an empty niche corresponds to
an initial condition of identical, rather than random, sequences.
The population dynamics again follows Eq.\ (\ref{2}).
To see an
effect of modularity, an expansion to 4th order in time is required.
As discussed in Appendix B, we find 
\begin{equation}
\langle f(t) \rangle = 2 L + b t^2 + c t^3 + d t^4
\label{6}
\end{equation}
with 
\begin{eqnarray}
 (2!) b &=& 16 \mu L (1-1/N)
\nonumber \\
 (3!) c &=& -64 \mu L^2 (1-1/N)/N - 192 \mu^2 L (1-1/N) 
- 32 \mu \nu L (1-1/N)/N 
\nonumber \\
(4!) d &=&
128 \mu^2 \nu L M (1-1/K) (1-1/N) (1-4/N )
+
64 \mu L (1-1/N) [
 2 (6 + 14 \mu^2 - \mu \nu) 
 \nonumber \\
&&+2 \mu \nu (1-4/N)  /K
-2 (36 - 10 \mu L - 9 \mu \nu)/N
+   (4 L^2 + 2 L + 92 + 4 \nu L + \nu^2)/N^2]
\nonumber \\
\label{101}
\end{eqnarray}
Interestingly, the fitness function initially increases quadratically.
The modularity dependence enters at fourth order, and
$\langle f_{M>0}(t) \rangle > \langle f_0(t) \rangle$ 
for short times.

\section{Comparison of Theory to Simulation Results}

\begin{figure}[tb!]
\begin{center}
\epsfig{file=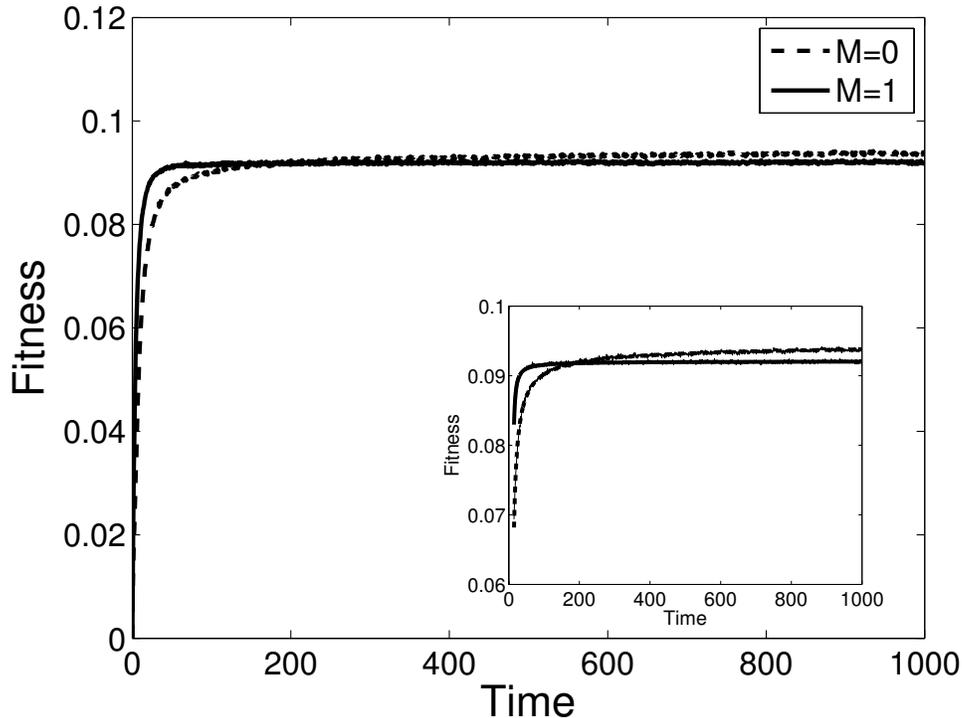,width=0.90\columnwidth,clip=}
\end{center}
\caption{Shown is the population average fitness for the modular
($M=1$, solid) and non-modular ($M=0$, dashed) systems. 
Fitness is normalized by $L$ and the offset $2 L$ is subtracted.
The modular system evolves to a greater fitness value
for times $t<t^*$.  Here $t^* \approx 182$.
Here Eq.\ (\ref{1}) has been scaled by $\epsilon=0.1$,
and $L=100$, $N = 10 000$, $\mu = 0.05$, $\nu = 0.6$, $K = 5$, and
$C = L /K-1$, motivated by application to evolution of
influenza to be discussed below.
In inset is shown the fit of the data to the form of Eq.\ (\ref{4a}).
}
\label{fig2}
\end{figure}

The response function of the modular and non-modular system can be
computed numerically as well.  
We start the system off with random initial sequences, so that
the average initial $\langle f(0) \rangle - 2L$ is zero,
and compute the
evolution of the average fitness, $\langle f(t) \rangle $.
In Fig.\ \ref{fig2} we show the results
from a Lebowitz-Gillespie simulation.  We see that 
$\langle f_M(t) \rangle > \langle f_0(t) \rangle $
for $t < t^*$ for $M=1$.  That is, the modular system evolves
more quickly to improve the average fitness, for times
less than a crossover time, $t^*$.  For $t> t^*$, the
constraint that modularity imposes on the connections
leads to the non-modular system dominating, 
$\langle f_0(t) \rangle > \langle f_M(t) \rangle $.

Eq.\ (\ref{3})  shows these results analytically, and
we have checked Eq.\ (\ref{3}) by numerical simulation as well.
For the parameter values of Fig.\ \ref{fig2}
and $M=0$,
theory predicts $a/L = 0.019998$, $b/L = -0.011636$, and
simulation results give 
$a/L = 0.020529 \pm 0.000173$, $b/L = -0.011306 \pm 0.000406$. 
For $M=1$,
theory predicts $a/L = 0.019998$, $b/L = -0.002041$, and
simulation results give 
$a/L - 0.019822 \pm 0.000200$, $b/L = -0.0019084 \pm 0.000177$

\begin{figure}[tb!]
\begin{center}
\epsfig{file=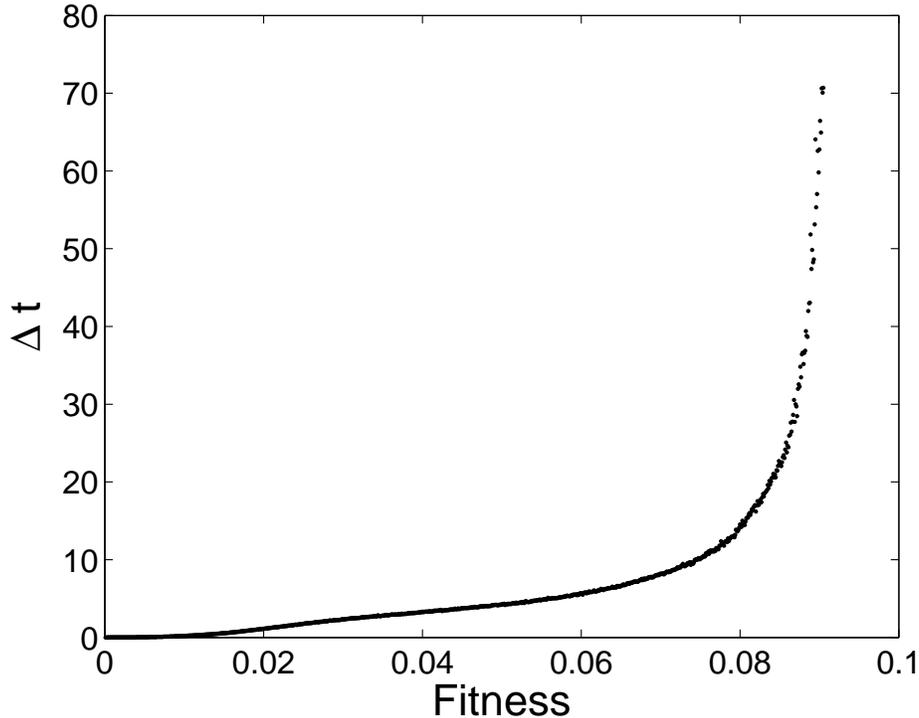,width=0.90\columnwidth,clip=}
\end{center}
\caption{Shown is the difference in time required, $\Delta t$, for 
the  modular ($M=1$) and  non-modular ($M=0$) system
to reach a specified average fitness value for times
$t < t^*$, starting
from random initial sequences.
Parameter values are as in Fig.\ \ref{fig2}.
}
\label{fig1}
\end{figure}

Since the landscape is rugged, we expect the time it takes the system
to reach a given fitness value, $\langle f(t) \rangle$, 
starting from random sequences to grow in a
stretched exponential way 
at large times.
Moreover, since the combination of horizontal gene transfer and
modularity improves the response function at short times, we expect
that the difference in times required to reach $\langle f(t) \rangle$
of a non-modular and modular system
also increases in a stretched exponential way, by analogy with statistical
mechanics, in which we expect the spin glass energy
response function to converge
as 
\begin{equation}
\langle f(t) \rangle \sim f_\infty - c \ln^{- 2/\nu} t/t_0 
\label{4a}
\end{equation}
with $\nu = 1$ \cite{Dotsenko}.
Figure \ref{fig2} shows the fit of the data to this functional
form, 
with $f_\infty = 0.958$, $c= 0.067$, $t_0  = 3.168$ for $M=0$
and
$f_\infty = 0.922$, $c= 0.004$, $t_0  = 7.619$ for $M=1$.
Figure \ref{fig1} shows the stretched exponential speedup in the rate
of evolution that modularity provides.

\begin{figure}[tb!]
\begin{center}
\epsfig{file=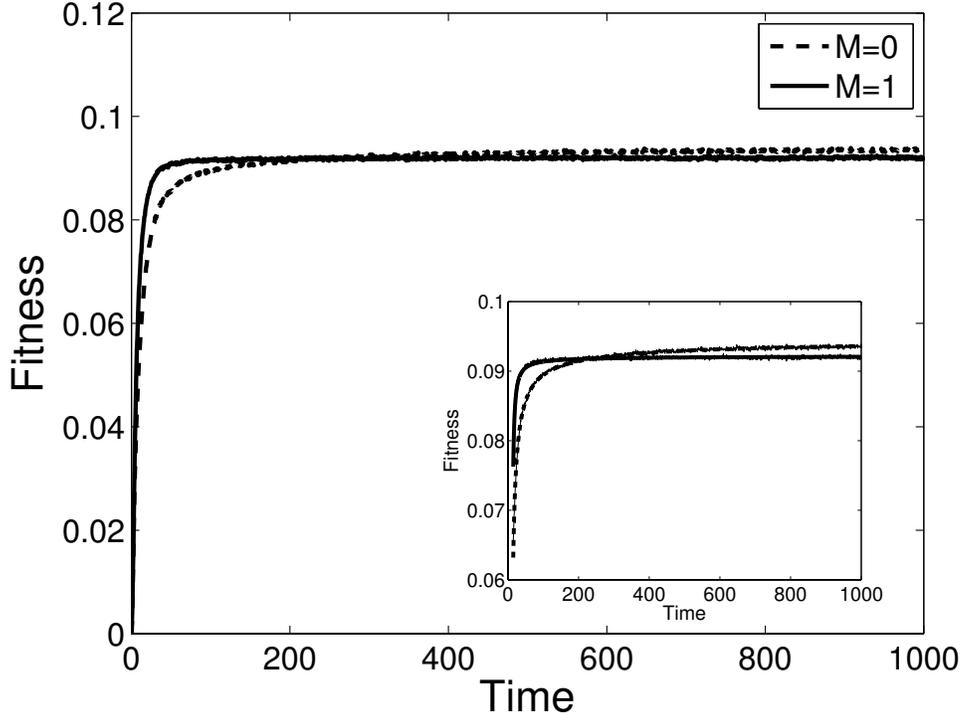,width=0.90\columnwidth,clip=}
\end{center}
\caption{Shown is the population average fitness for the modular
($M=1$, solid) and non-modular ($M=0$, dashed) systems
when all individuals initially have the same sequence.
Fitness is normalized by $L$ and the offset $2 L$ is subtracted.
The modular system evolves to a greater fitness value
for times $t<t^*$.  Here $t^* \approx 230$.
Parameter values are as in Fig.\ \ref{fig2}.
\label{fig3}
}
\end{figure}

The prediction for evolution of a population of identical sequences
in a new niche is shown in Eq.\ (\ref{101}). 
Analytically, the effect of modularity shows up at 4th order
rather than 2nd order when all sequences are initially identical.
Qualitatively, however, at all but the very shortest times, the results 
for random and identical sequences are similar, as shown in
Figs.\ \ref{fig2} and \ref{fig3}.

\section{Average Fitness in a Changing Environment}

We show how to use these results to calculate the average
fitness in a changing environment.  We consider that every $T$
time steps, the environment randomly changes.  During such an
environmental change, each $J_{ij}$ in the 
fitness, Eq.\ (\ref{1}), is randomly redrawn from the Gaussian
distribution with probability $p$.  That is, on average,
a fraction $p$ of the fitness is randomized.  Due to this
randomization, the fitness immediately after the environmental
change will be $1-p$ times the value immediately before the change,
on average.  This condition allows us to calculate the
average time-dependent fitness during evolution in one
environment as a function of $p$ and $T$, given only the
average fitness starting from random initial conditions,
$\langle f(t) \rangle$ \cite{Park13}.
We denote the average fitness reached during 
evolution in one environment as
$f_{p,T}(M)$.  This is related to the average fitness
with random initial conditions by
$f_{p,T}(M)= \langle f(M) \rangle (t)$,
 where $t$ is chosen
to satisfy
$ \langle f(M) \rangle (t-T)= (1-p) \langle f(M) \rangle (t)$
due to the above condition.
Thus, for high rates or large magnitudes of environmental
change, Eq.\ (\ref{3}) can be directly used along with these
two conditions to predict the steady-state average
population fitness.

\begin{figure}[tb!]
\begin{center}
a) \epsfig{file=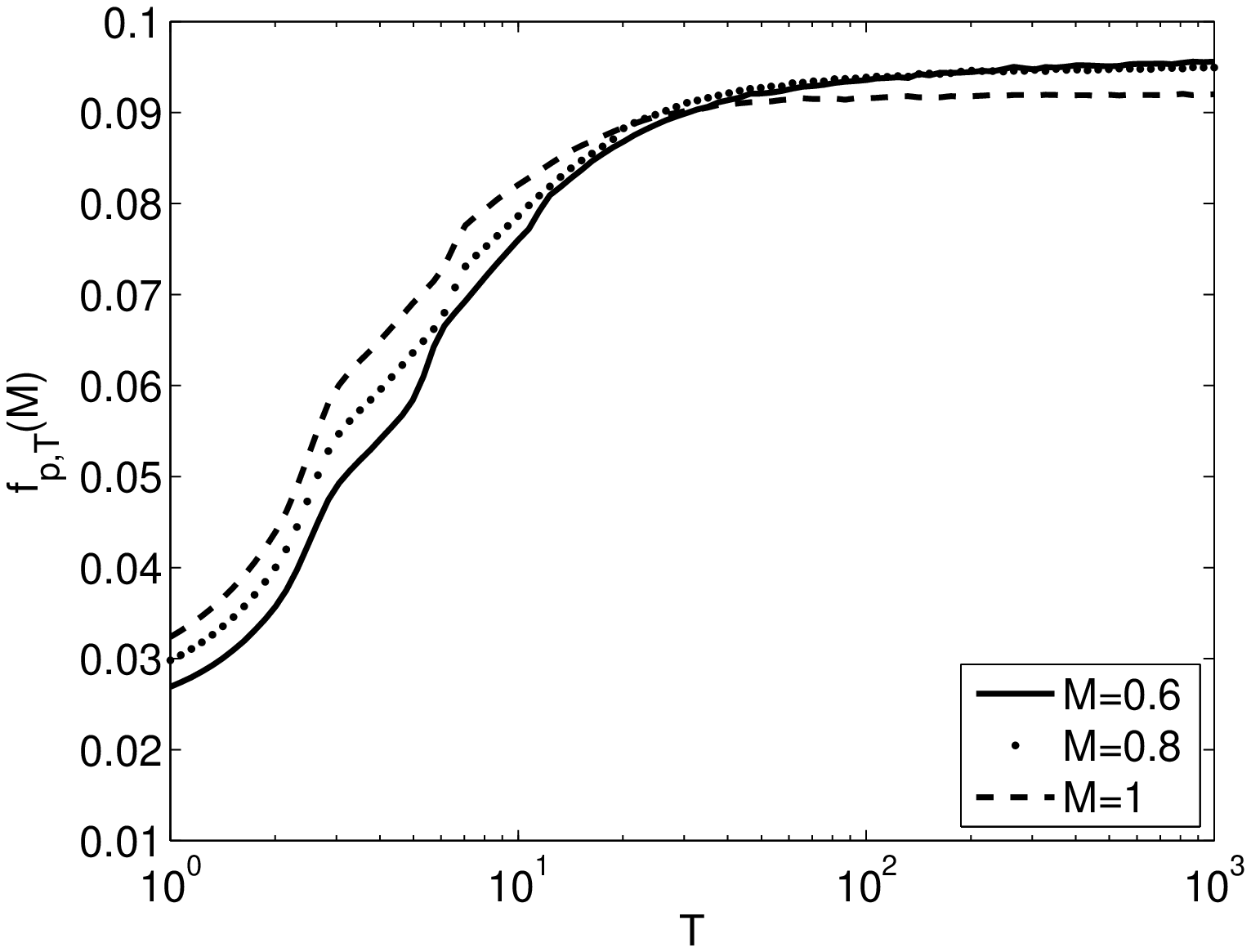,width=0.40\columnwidth,clip=}
b) \epsfig{file=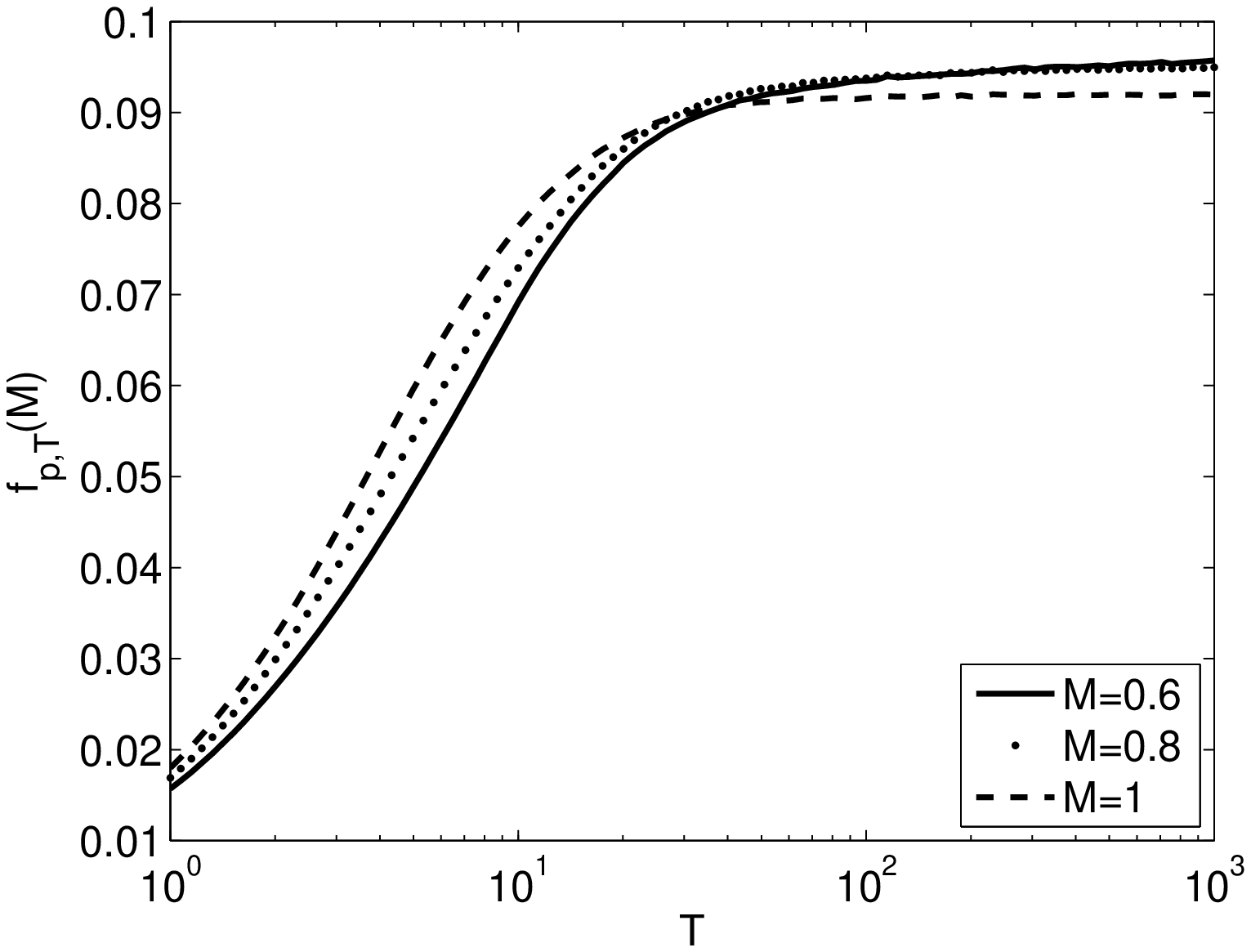,width=0.40\columnwidth,clip=}\\
c) \epsfig{file=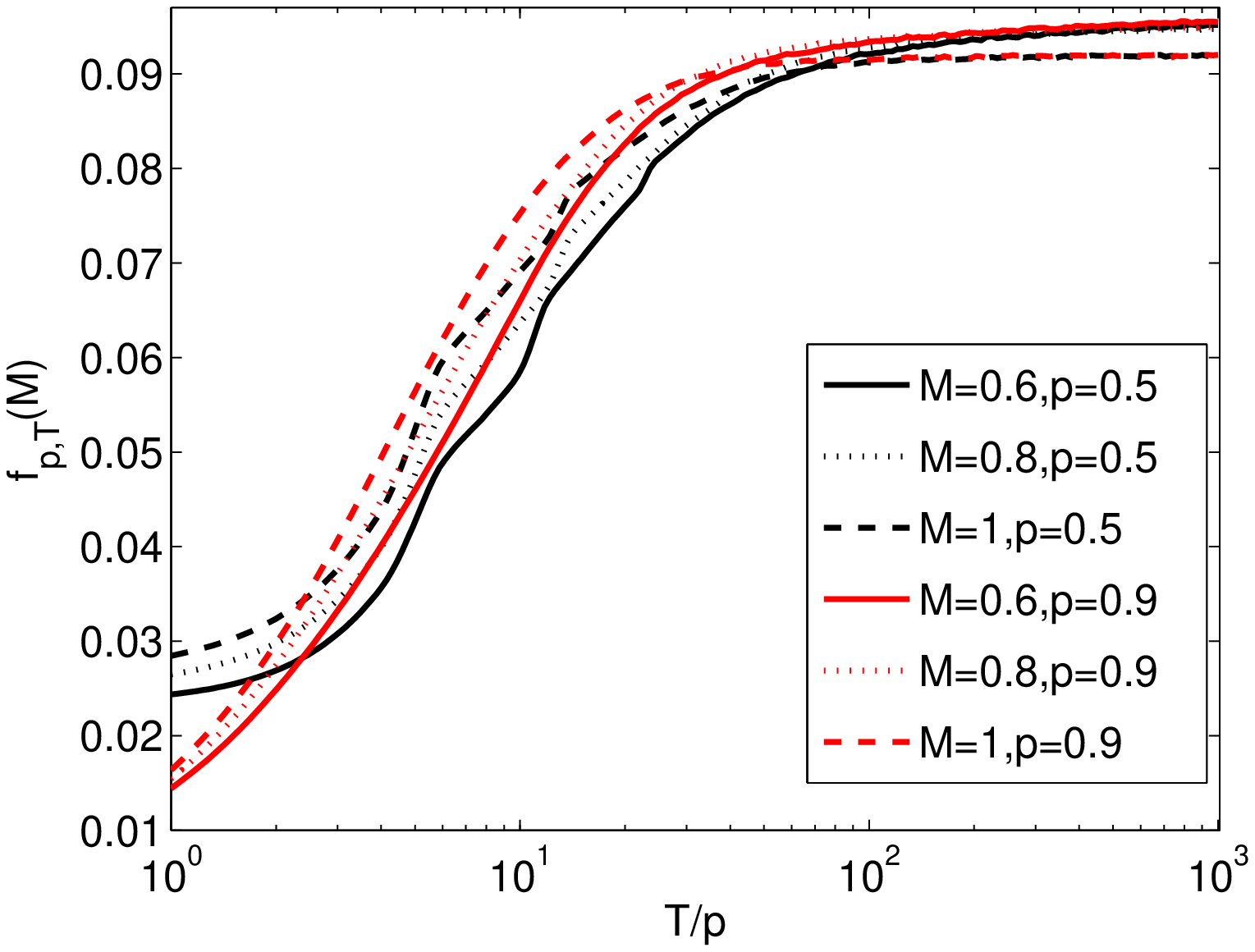,width=0.40\columnwidth,clip=}
\end{center}
\caption{
Shown is the fitness, $f_{p,T}(M)$, in an environment that changes
with characteristic time $T$ and magnitude 
a) $p=0.5$ or b) $p=0.9$. 
c) The fitness is replotted as a function of T/p.
For each $p$, the fitness is shown for three values of modularity: 
$M=0.6$ (solid), $M=0.8$ (dotted), and $M=1$ (dashed).
Parameters are as in Fig.\ \ref{fig2}.
\label{fig5}
}
\end{figure}
Figure  \ref{fig5} shows how the average fitness depends on the
rate and magnitude of environmental change.  The
average fitness values for a given modularity are computed numerically.
The figure displays a crossing of the 
modularity-dependent fitness curves.
The curve associated with the largest modularity is greatest at short
times.  There is a crossing to the curve associated with the second
largest modularity at intermediate times.  And there is a
crossing to the curve associated with the smallest modularity
at somewhat longer times.  In other words, the value of modularity
that leads to the highest average fitness depends on time, $T$, decreasing
with increasing time.

Figure  \ref{fig5} demonstrates
that larger values of modularity, greater $M$,  lead
to higher average fitness values 
for faster rates of environmental change, smaller $T$.
The advantage of greater modularity persists to larger
$T$ for 
greater magnitudes of environmental change, larger $p$.
In other words, modularity leads to greater average fitness
values either for greater frequencies of environmental change, $1/T$,
or greater magnitudes of environmental change, $p$.

It has been argued that an approximate measure of the environmental
pressure is given by the product of the magnitude and frequency
of environmental change $p/T$ \cite{Park13}.
If this is the case, then the curves in 
Fig.\  \ref{fig5}ab as a function of $T$ for different values of
$p$ should collapse to a single curve as a function of $T/p$.
As shown in Fig.\  \ref{fig5}c, this is approximately true
for the three cases $M=0.6$, $M = 0.8$, and $M=1$.

\section{Application of Theory to Influenza Evolution Data}


We here apply our theory to the evolution of the influenza virus.
Influenza is an RNA virus of the Orthomyxoviridae family.
The 11 proteins of the virus are encoded by 8 gene segments.
Reassortment of these gene segments is common \cite{swine,Holmes,Marshall}.
In addition, there is a constant source of genetic diveristy,
as most human influenza viruses arise from birds, mix in pigs,
with a select few transmitted to humans
\cite{swine,Holmes,Marshall,Trifonov}.
We model a simplification  of
this complex coevolutionary dynamics with
the theory presented here. 
We consider $K \approx 5$ modules, with mutations in the
genetic material encoding them leading to an
effective mutation rate at the coarse-grained length scale.
We do not consider individual amino acids,
but rather a coarse-grained $L=100$ reprsentation
the viral protein material.
The virus is under strong selective pressure
to adapt to the human host, having most often arisn in
birds and trasmitted through swine.  The virus is also under
strong pressure from the human immune system.
The virus evolves in response to this pressure, thereby increasing
its fitness.  The increase in fitness was estimated
by tracking the increase in frequency of each viral strain,
as observed in the public sequence databases:
$\ln [f_i(t+1) / f_i(t)] = \hat x_i(t+1) / x_i(t)$, where
$f_i(t)$ is the fitness of clade $i$ at time $t$,
$x_i(t)$ is the frequency of clade $i$ among all clades
observed at time $t$, and $\hat x_i(t+1)$ is  the
frequency at $t+1$ predicted by a model that includes
a description of the mutational processes
 \cite{Lassig}.  Here ``clade'' is 
a term for the quasispecies of closely-related influenza
sequences at time $t$.   We use these approximate fitness values,
estimated from observed HA sequence patterns and
an approximate point mutation model of evolution, for comparison to the
present model.

\begin{figure}[tb!]
\begin{center}
\epsfig{file=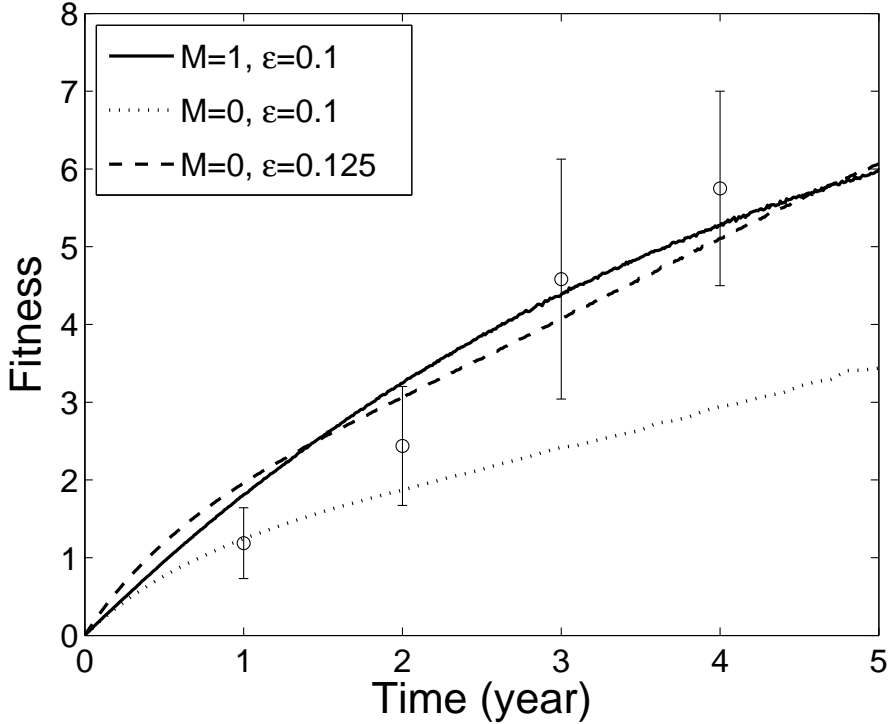,width=0.90\columnwidth,clip=}
\end{center}
\caption{Shown is the fitness of influenza virus H3N2 in humans,
estimated from the time derivative of the frequency with which sequences
appear in the GenBank sequence database.
Each time a new circulating strain is introduced to humans, typically
from pigs infected from birds, the fitness increases
as a quasispecies expands around the strain.  
The average fitness increase over
four such antigenic shifts between 1993 and 2010 is shown.
Data are from \cite{Lassig}.
Error bars are one standard error.
Theoretical results are shown for the modular ($M=1$, solid) and
non-modular ($M=0$, dotted) model predictions.
Theory was fit to the average linear fitness increase at 5 years.
Parameter values are as in Fig.\ \ref{fig2}, with fitness
scaled by $\epsilon=0.1$.
Theoretical results are also shown for $\epsilon = 0.125$ and
$M=0$ (dashed).
}
\label{fig4}
\end{figure}

Influenza evolves within a cluster of closely-related
sequences for 3--5 years and
then jumps to a new cluster \cite{Smith2,Gupta,He}. 
Indeed, the fitness flux data over 17
years \cite{Lassig} shows a pattern of discontinuous changes
every 3--5 years. 
Often these jumps are related to influx of genetic
material from swine \cite{swine,Holmes,Marshall,Trifonov}.
By clustering the strains, the clusters and the transitions
between them can be identified:
the flu strain evolved from Wuhan/359/95 to Sydney/5/97 at 1996-1997,
Panama/2007/1999 to Fujian/411/2002 at 2001-2002, California/7/2004
to Wisconsin/67/2005 at 2006-2007, Brisbane/10/2007 to
BritishColumbia/RV1222/09 at 2009-2010 \cite{He}. 
These cluster transitions correspond to the
discontinuous jumps in the fitness flux evolution and
discontinuous changes in the sequences.
Our theory represents the evolution within one
of these clusters, considering reassortment
only among human viruses.
Thus, we predict the evolution of the fitness during each of these
periods.  There are 4 periods within the time frame 1993--2009.
Figure \ref{fig4} shows the measured fitness data \cite{Lassig}, averaged
over these four time periods.

We scaled Eq.\ (\ref{1}) by $\epsilon$ to fit the observed data.
The predictions from Eq.\ (\ref{3}) are shown in Fig.\ \ref{fig4}.
We find $\epsilon = 0.1$ and $M=1$ fit the observed data well.
We assume the overall rate of evolution is equally contributed to
by mutation and horizontal gene transfer. The average observed
substitution rate in the 100aa epitope region of the HA protein is
5 amino acids per year \cite{He,Lassig}.  We interpret this
result in our coarse-grained model to imply
$\mu = .05$ and  $\nu = 0.6$. 

The value of $\epsilon$ required to fit the non-modular model to the
data is 25\% greater than the value required to fit the modular
model to the data.    
That is, approximately 25\% of the 
observed rate of fitness increase of the virus
could be ascribed to a modular viral fitness landscape.
Thus, to achieve the observed rate of evolution,
the virus may either have evolved modularity $M=1$, or the virus
may have evolved a 25\% increase in its replication rate by
other evolutionary mechanisms.
Given the modular structure of the epitopes on the 
haemagglutinin  protein of the virus, the modular nature of the
viral segments, and the modular nature of naive, B cell, and T
cell immune responses we suggest that influenza
has likely evolved a modular fitness landscape.

\section{Generalization to Investigate the Effect of Landscape Ruggedness}

We here generalize Eq.\ (\ref{1}) to $p$-spin interactions, $q$ letters in 
the alphabet from which the sequences are constructed, and what is known
as the GNK random matrix form of interactions \cite{deem2003sequence}. These generalizations
allow us to investigate the effect of landscape ruggedness on the
rate of evolution.

\subsection{Generalization to the $p$-spin SK Model}
We first generalize the SK model to interactions among $p$ spins, which
we term the $p$SK model. The $p$SK landscape is more rugged for increasing  $p$.
Thus, we might expect modularity to play a more important role
for the models with larger $p$.
The generalized fitness is written as
\begin{equation}\label{eq:manybody}
f=\sum_{i_1i_2...i_p}J_{i_1i_2...i_p}\Delta_{i_1i_2...i_p}s_{i_1}s_{i_2}...s_{i_p}+2L
\end{equation}
The $J$ and $\Delta$ tensors are symmetric, that is, $J_{i_1i_2...i_p}=J_{[i_1i_2...i_p]}$ and $\Delta_{i_1i_2...i_p}=\Delta_{[i_1i_2...i_p]}$, where $[i_1i_2...i_p]$ is any permutation of $i_1i_2...i_p$. 
The probability of a connection is $C_1/L^{p-1}$  inside the block, $\lfloor i_1/l \rfloor = \lfloor  i_2/l \rfloor=...=\lfloor  i_p/l \rfloor$,
and $C_0/L^{p-1}$ outside the blocks. The connections are defined as
\begin{equation}
\sum_{i_2...i_p}\Delta_{i_1i_2...i_p}=C
\end{equation}
Since $C=C_0[1-(\frac{1}{K})^{p-1}]+C_1(\frac{1}{K})^{p-1}$, it follows
that $C_1=C[MK^{p-1}+(1-M)]$, and $C_0=C(1-M)$.

Increasing the value of $p$
makes the landscape more rugged, and so it is more difficult for system to
reach a given value of the fitness in a finite time.
Thus, modularity is expected to play a more significant role in
giving the system an evolutionary advantage. Specifically we expect 
that the effect of modularity will show up at shorter times for
larger $p$.  As shown in Appendix B, we find
\begin{eqnarray}
\nonumber a&=&2L(1-1/N)
\\\nonumber 2b&=&-\frac{8L^2}{N}(1-1/N)-4\mu pL(1-1/N)
\\&&+2\nu
L\left[\left(\frac{2}{N}-p+\frac{3p}{N}\right)(1-M)\left(1-\frac{1}{K^{p-1}}\right)
-\frac{2}{N}\right](1-1/N)
\end{eqnarray}

\begin{figure}[tb!]
\begin{center}
\includegraphics[width=0.8\textwidth]{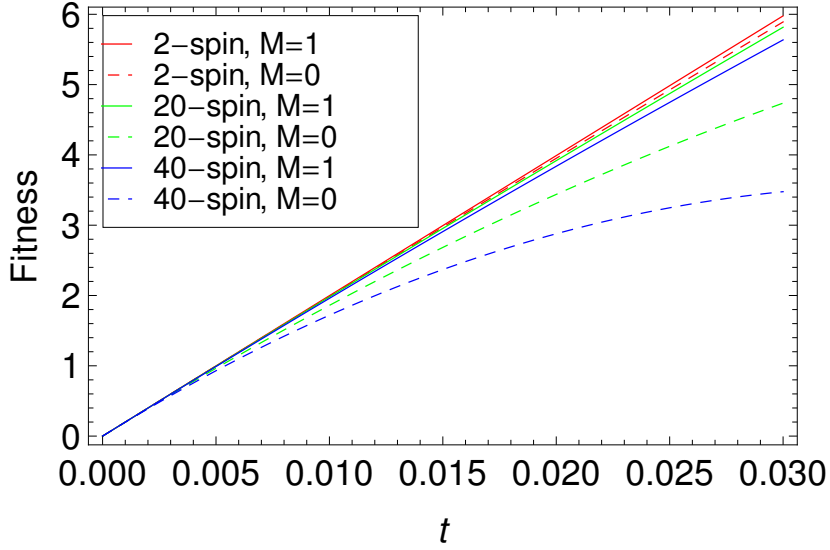}
\end{center}
\caption{Results for $2$-spin, $12$-spin and $30$-spin SK interaction fitness when
initial sequences are random. The fitness is given by Eq.\ 
(\ref{eq:manybody}). 
The parameters are as in Fig.\ \ref{fig2}.
\label{fig:p-body}
}
\end{figure}

The modular terms increase faster with $p$ than the non-modular terms,
 so the dependence on modularity of evolution is more significant with 
bigger $p$. 
Figure \ref{fig:p-body} shows the average fitness curves
for the $p$-spin interaction. 
Due to the normalization of the $J$ values, all fitness curves
have the same slope at $t=0$, i.e.\ the same values of $a$.
However, increasing  $p$ leads to slower rates of fitness increase
at finite time, especially for the $M=0$ case.  At finite time, for large
values of $p$, the evolutionary advantage of larger $M$ is more
pronounced.  That is, 
modularity help the sequences to evolve more efficiently.

We also calculated the average fitness for the initial condition of
all sequences identical.  As discussed above, many-spin effects
make the landscape more rugged for larger $p$.
We seek to understand this effect quantitatively
for the case of same initial sequences as well.
As shown in Appendix B, the Taylor expansion results are
\begin{eqnarray}
\nonumber (2!) b&=&8\mu pL(1-1/N)
\\\nonumber (3!) c&=& -32\mu pL^2(1-1/N)/N-16\mu^2p^2L(1-1/N)
\\&&\nonumber-64\mu^2pL(1-1/N)-16\mu\nu pL(1-1/N)/N
\\\nonumber (4!) d&=&64 \mu^2 \nu p(p-1)L M (1-1/K) (1-1/N) (1-4/N )
\nonumber \\ &&
+
32 \mu pL (1-1/N) [
 (6p + 7 \mu^2 p^2- 2\mu \nu (p-1)) 
 \nonumber \\
&&+2 \mu \nu (1-4/N)  /K
-(36p - 10 \mu pL - 5 \mu \nu p-8\mu\nu (p-1))/N
\nonumber \\ &&
+  (4 L^2 + 2 L + 46p + 4 \nu L + \nu^2)/N^2]\nonumber
\\
\end{eqnarray}

\subsection{Generalization to the Planar Potts Model}

We here generalize the alphabet from two-states, $s_i = \pm 1$, to 
$q$ states, with a 
planar Potts model  \cite{wu1982potts} of the fitness. 
The $q=2$ model is often justified as a projection of the $q=4$ nucleic
acid alphabet onto purines and pyrimidines.  Thus, $q=4$ is a
relevant generalization.  Additional $q=20$ corresponds to 
the amino acid alphabet.  The alphabet size can affect the evolutionary
phase diagram \cite{Munoz}, so
$q$ is a relevant order parameter.
In the planar Potts model, each vector spin takes on $q$ equidistant
angles, and the angle between two is defined by
$\boldsymbol{s}_i\cdot \boldsymbol{s}_j=\cos \theta_{ij}$.
 The fitness is
\begin{equation}\label{eq:potts}
f=\sum_{ij}J_{ij}\Delta_{ij}\cos\theta_{ij}+2L
\end{equation}
The directions of a spin are evenly spaced in the plane, so when the spin points to $q$ directions, the angle between two spins is $2k\pi/q$, where $k$ can be $0$, $1$, ..., $q-1$. 

\begin{figure}[tb!]
\begin{center}
\includegraphics[width=0.8\textwidth]{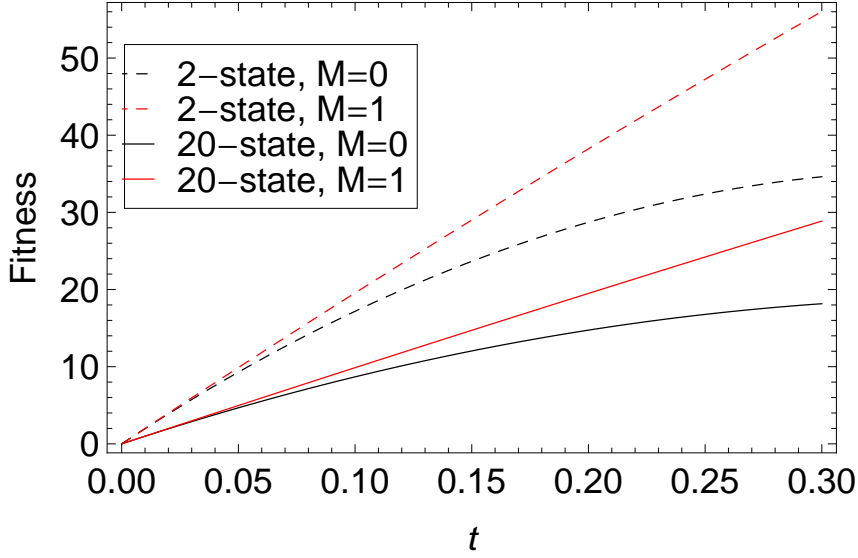}
\end{center}
\caption{Average fitness in the planar Potts model
for alphabet sizes of $q=2$ or $q=20$.
Sequences are initially random. The fitness Eq.\  (\ref{eq:potts}). 
Other parameters are same as in Fig.\ \ref{fig2}.
\label{fig:p-spin}
}
\end{figure}

The evolutionary dynamics in the planar Potts model is distinct from that
in the $p$SK model.  As shown in Appendix B,
\begin{eqnarray}
\nonumber a&=&L(1-1/N)(1+\delta_{q,2})
\\\nonumber2b&=&-4L^2(1-1/N)(1+\delta_{q,2})/N-2\mu L(1-1/N)\frac{q}{q-1}(1+\delta_{q,2})\nonumber
\nonumber \\ && 
- 2 \nu L \left[
\left( 1-\frac{1}{K}\right)
\left( 1-M \right)
\left( 1-\frac{4}{N}\right)
+
\frac{1}{N}
\right]
\left( 1-\frac{1}{N}\right) 
(1+\delta_{q,2})
\end{eqnarray}
Figure \ref{fig:p-spin} displays results for two values of $q$.
Increasing $q$ does 
increase the landscape ruggedness in the planar Potts model.
There is an evolutionary speedup provided by modularity.
The $q$ dependence, however, is not as dramatic as in the
$p$SK model.

\subsection{Generalization to the GNK Model}
A model related to the SK form is the GNK model, 
a simple form of which is
\begin{equation}
f=H+2L=\sum_{ij}\sigma_{ij}(s_i,s_j)\Delta_{ij}+2L
\end{equation}
where the $\sigma$ matrix is a symmetric matrix with random Gaussian entries, and the $\Delta$ matrix is the same as that in Eq.\ (\ref{1}). 
Other than the condition of symmetry, the entries in the $\sigma$ matrices are
independently drawn from a Gaussian distribution of zero mean and unit variance
in each matrix and for different $i,j$.  Thus, other than the condition
$\sigma_{ij} = \sigma{ji}$, the values
are independent for different $i$, $j$, $s_i$, or $s_j$. 
This form is generalized to a $p$-spin, $q$-state form as
\begin{equation}\label{eq:gnk}
f=\sum_{i_1i_2...i_p}\sigma_{i_1i_2...i_p}(s_{i_1},s_{i_2},...,s_{i_p})\Delta_{i_1i_2...i_p}+2L
\end{equation}
where $s_i$ can take one of the $q$ values.

For the GNK model, as shown in Appendix B, $p$-spin $q$-state result is
\begin{eqnarray}
\nonumber a&=&2L(1-1/N)(1-1/q^p)/\ln q
\\\nonumber 2b&=&-\frac{8L^2}{N}(1-1/N)(1-1/q^p)/\ln q-2\mu pL(1-1/N)\frac{q}{(q-1)\ln q}
\\&&-2\nu KL(1-\frac{1}{N})(1-M)\Bigg[(1-\frac{3}{N})
\left(1+\frac{1}{q^p}-\left(\frac{1-1/K+q/K}{q}
\right)^p-\left(\frac{q(1-1/K)+1/K}{q}\right)^p\right)\nonumber
\\&&-2\left(\frac{1-1/K+q/K}{q}\right)^p/N
+2/NK+2(1-1/K)/Nq^p\Bigg]/\ln q\nonumber
\\&&+4\nu L(1-\frac{1}{N})\left(1+(K-1)/q^p\right)/(N\ln q)
\end{eqnarray}
As with previous models, modularity increases the rate of evolution.

\section{Conclusion}
The modular spin glass model captures several important aspects
of biological evolution: finite population, rugged fitness landscape,
modular correlations in the interactions, and horizontal
gene transfer.  We showed in both numerical
simulation and analytical calculations that a modular landscape allows
the system to evolve higher fitness values for times $t < t^*$.
Using this theory to analyze fitness data extracted from influenza
virus evolution, we find that 
approximately 25\% of the observed rate of fitness increase of the virus
could be ascribed to a modular viral landscape.
This result is consistent with the
success of the modular theory of viral immune recognition, termed 
the $p_{\rm epitope}$ theory, over the non-modular theory, termed
$p_{\rm sequence}$  \cite{Gupta}.  The former correlates with human
influenza vaccine effectiveness with $R^2 = 0.81$, whereas the latter
correlates with $R^2 = 0.59$.
The model, in general, analytically demonstrates a
selective pressure for the prevalence of modularity in biology.
The present model may be useful for understanding the influence of
modularity on other evolving biological systems, for example the HIV
virus or immune system cells.

\section*{Acknowledgments}
This research was supported
by the US National Institutes of Health under grant
1 R01 GM 100468--01.
JMP was also supported by the National Research Foundation of Korea
Grant (NRF-2013R1A1A2006983).

\section{Appendix A: The Requirement that Fitness be Non-Negative}

\subsection{The $p$SK Model}\label{sec:norm}
The normalization of the
coupling in Eq.\ (\ref{1}), the $J_{ij}$, affects the
maximum and minimum possible values of the $H$ term.
We require that the average energy per site of the sequence be finite when $L\rightarrow \infty$, i.e.\ that $H$ is on the order of $L$. For the $2$-spin, $2$-state interaction, the Wigner semicircle law \cite{wigner1958distribution} shows that a rank-$L$ random symmetric matrix with $\sum_j \langle J_{ij}^2 \Delta_{ij}^2 \rangle=1 ~\forall~ i$ has a minimum eigenvalue $-2$. Thus, if we consider the $s_i$ in Eq.\ (\ref{1}) to be normalized as $\sum_is_i^2=L$, the $2L$ shift in Eq.\  (\ref{1}) guarantees that $H+2L\ge 0$. So after considering the connection matrix $\Delta$, 
we choose $J_{ii} = 0$ and
 $\langle J_{ij} J_{kl} \rangle=(\delta_{ik} \delta_{jl}+
 \delta_{il} \delta_{jk})
 /C$.

For the  $p$SK case, we first assume that $J_{i_1...i_p}$ is not symmetric, i.e. $J_{i_1...i_p}$ with permutations of the same labels are drawn independently.
We will consider the symmetrization afterward.
We define
\begin{equation}
K_{i_1i_2}=\sum_{i_3...i_p}J_{i_1i_2...i_p}\Delta_{i_1i_2...i_p}
\end{equation}
And for any $i_1$
\begin{equation}
\left<\sum_{i_2}K_{i_1i_2}^2\right>=\sum_{i_2...i_p}\langle J_{i_1i_2...i_p}^2\rangle\Delta_{i_1i_2...i_p}=C\langle J^2\rangle
\end{equation}
So,
\begin{eqnarray}
\nonumber H&=&\sum_{i_1i_2...i_p}s_{i_1}s_{i_2}...s_{i_p}J_{i_1i_2...i_p}\Delta_{i_1i_2...i_p}
\\\nonumber &=&\sum_{i_1i_2}s_{i_1}s_{i_2}(\sum_{i_3...i_p}s_{i_3}...s_{i_p}J_{i_1i_2...i_p}\Delta_{i_1i_2...i_p})
\\\nonumber &=&s_{i_1}s_{i_2}K_{i_{1}i_{2}}
\nonumber\\&=&s_{i_1}s_{i_2}\frac{K_{i_{1}i_{2}}+K_{i_{2}i_{1}}}{2}
\end{eqnarray}
In the last step we symmetrize the $K$ matrix so that we can apply the Wigner semicircle law. From the semicircle law, if we want the minimum of the $p$-spin interaction to be $-2L$, we need $\sum_{i_2}\langle(K_{i_1i_2}+K_{i_2i_1})/2\rangle^2=1$ for every $i_1$.  We, thus, find
\begin{equation}
\sum_{i_2}\langle\frac{K_{i_{1}i_{2}}+K_{i_{2}i_{1}}}{2}\rangle^2=\frac{1}{4}\sum_{i_2}\langle K_{i_1i_2}^2+K_{i_2i_1}^2\rangle=\frac{C}{2}\langle J^2\rangle=1
\end{equation}
So $\langle J_{i_1...i_p}^2\rangle=2/C$ for asymmetric $J_{i_1...i_p}$. We symmetrize the $J_{i_1...i_p}$
\begin{equation}
J'_{i_1i_2...i_p}=\frac{1}{p!}\sum_{\text{permutations}}J_{i_1i_2...i_p}
\end{equation}
to find
\begin{equation}
\langle J'^{2}_{i_1i_2...i_p}\rangle=\frac{p!}{p!^2}\langle J_{i_1i_2...i_p}^2\rangle=\frac{2}{p!C}
\end{equation}

The Wigner approach to calculate the minimal possible value of $H$ is overly
conservative, because in this approach all possible real vectors, $s_i$
are considered, whereas in our application $s_i = \pm 1$.
We here use extreme value theory to take this constraint
into account. We use the $2$-spin interaction to exemplify the method. As the $J$ matrix is symmetric, its elements are not independent, so the terms in $H$ are not independent.
We rewrite $H$, noting the diagonal terms are zero, as
\begin{equation}
H=\sum_{ij}J_{ij}s_is_j\Delta_{ij}=2\sum_{i<j}J_{ij}s_is_j\Delta_{ij}=2H'
\end{equation}
The terms in $H'$ are independent. After taking into account the connection matrix, there are $LC/2$ non-zero independent elements. As $s_i$ is either $+1$ or $-1$, multiplying them does not change the distribution of each element, so the $H'$ is the sum of $LC/2$ Gaussian variables. We choose the variance of each element as $1/C$, so $H'\sim N(0,L/2)$, and $H$ is twice of this, so 
\begin{equation}
H\sim N(0,2L)
\end{equation}

There are $2^L$ different sequences, so there are $2^L$ different $H$, 
which differ
by a few terms instead of all terms, so they are not independent.
We seek the smallest $H$.
 Slepian's lemma
\cite{li2001gaussian} shows that for Gaussian variables $X_i$ and $Y_i$ with average $0$, if $\langle X_i^2\rangle=\langle Y_i^2\rangle$, and $\langle X_iX_j\rangle\leq\langle Y_iY_j\rangle$ for all $i$ and $j$, then for any $x$, 
\begin{equation}\label{eq:slepian}
P(\max_{1\leq i\leq n}X_i\leq x)\leq P(\max_{1\leq i\leq n}Y_i\leq x)
\end{equation}
Thus, since $X_i$ and $Y_i$ have zero mean, the minimum of the less correlated variables is smaller than that of the more correlated ones. Thus, we still use extreme value theory to determine the minimum for the same number of independent variables with the same distribution, and we use this number as an estimate of a lower bound. When we add this estimated value to $H$,  Slepian's lemma guarantees the non-negativity of the fitness. We will show shortly that this estimated value is quite near the true value. 
The extreme value theory calculation proceeds as follows:
\begin{equation}\label{eq:extreme}
\int_{-\infty}^{H_{\text{min}}}P(H)dH=1/2^L
\end{equation}
where $2^L$ is the sample space size when the spin has 2 directions. From the symmetry of Gaussian distribution and $H_{\text{max}}=-H_{\text{min}}$, we find
\begin{eqnarray}
\nonumber\frac{1}{\sqrt{2\pi}\sqrt{2L}}\int_{-\infty}^{H_{\text{max}}}e^{-x^2/4L}dx&=&0.5+\frac{1}{\sqrt{\pi}}\int_{0}^{H_{\text{max}}/2\sqrt{L}}e^{-x^2}dx
\\\nonumber&=&0.5+0.5\erf(H_{\text{max}}/2\sqrt{L})
\\\nonumber&\approx&1-\frac{e^{-H_{\text{max}}^2/4L}}{\sqrt{\pi}H_{\text{max}}/2\sqrt{L}}
\\&=&1-\frac{1}{2^L}
\end{eqnarray}
where $\erf()$ is the error function.
Since $L$ is large, we find
\begin{equation}
\frac{e^{-H_{\text{max}}^2/4L}}{\sqrt{\pi}H_{\text{max}}/2\sqrt{L}}\approx e^{-H_{\text{max}}^2/4L}=e^{-H_{\text{min}}^2/4L}=2^{-L}
\end{equation}
So
\begin{equation}\label{eq:min}
H_{\text{min}}=-2L\sqrt{\ln 2}\approx-2\times0.832L
\end{equation}

Numerical results show that the exact value is 
$-2\times 0.763L$ \cite{PhysRevB.76.184412}, quite close to 
the value in Eq.\ (\ref{eq:min}).
In this way, for the $p$SK interaction, 
a normalization  of
\begin{equation}
\langle J_{i_1,i_2,...,i_p}^2\rangle=2/Cp!
\end{equation}
gives a minimum $\ge -2L\sqrt{\ln 2}$. 
This bound becomes exact as $p\rightarrow\infty$ \cite{0305-4470-30-24-010}. As $p$ becomes bigger, the landscape becomes more rugged, and the sequences becomes more uncorrelated.
The correlation of general $p$ falls between that of $p=2$ and $p\rightarrow\infty$, so according to Slepian's lemma, the exact minimum of any $p$ will fall between that of $p=2$ and that of $p\rightarrow\infty$. As $0.832\approx 1$ and $0.763\approx 1$, we will neglect this factor, and set the shift to $2L$ when $\langle J^2\rangle=2/p!C$, which guarantees $f > 0$.
 
Here we also calculate $\langle H^2\rangle$, which is used to obtain the Taylor expansion results in section \ref{sec:details}. For a two-spin interaction we find
\begin{eqnarray}
\nonumber\langle H^2\rangle &=&\left<\sum_{ij}\sum_{kl}J_{ij}J_{kl}\Delta_{ij}\Delta_{kl}s_is_js_ks_l\right>
\\\nonumber&=&\sum_{ij}\sum_{kl}(\delta_{ik}\delta_{jl}/C+\delta_{il}\delta_{jk}/C)\Delta_{ij}\Delta_{kl}s_is_js_ks_l
\\&=&2\sum_{ij}\Delta_{ij}/C=2L
\end{eqnarray}
We similarly calculate $p$-spin results.  We find that $\langle H^2\rangle$ is always $2L$ for all $p$ under this normalization scheme.

\subsection{The Planar Potts Model}
For the planar Potts model, we consider that $q$ is even, so the configuration space of $q=2$ is a subset of that of $q>2$, and the minimum of $q>2$ is no bigger than that of $q=2$ Potts model, which is $-0.763 \times 2 L$ \cite{PhysRevB.76.184412}. 
We consider two limiting cases to obtain the lower bound of the minimum. First, when $q\rightarrow\infty$, the planar Potts model becomes the XY model, which is defined in this case as $H=\sum_{ij}J_{ij}\boldsymbol{s}_i\cdot\boldsymbol{s}_j=\sum_{ij}J_{ij}\cos\theta_{ij}$, where $\theta_{ij}$ is the angle between vector $\boldsymbol{s}_i$ and $\boldsymbol{s}_j$, and these vectors $\boldsymbol{s}_i$ can point to any direction in a two dimensional space. So the configuration space of the planar Potts model is a subset of that of the XY model, and the minimum of planar Potts model is no smaller than that of the XY model. 
Numerical results show that the ground state energy for the XY model is roughly $0.90\times 2L$ \cite{grest1983ground}.

Second, the vectors in the XY model are restricted in a two dimensional space. If we generalize it to an $n$ dimensional space, with $\boldsymbol{s}_i=(x_1,x_2,...,x_n)$, and $x_1^2+x_2^2+...+x_n^2=1$, we obtain the $n$ vector model. When $n\rightarrow\infty$, the model becomes the spherical model, the exact minimum of $H$ can be calculated analytically as $-2L$ when $\langle J_{ij}^2\rangle=1/L$  \cite{PhysRevLett.36.1217}. Similarly, we see that the configuration space of the XY model is a subset of the spherical model, and the minimum of the XY model is no smaller than the spherical model. So considering the above two limiting cases, the minimum of the planar Potts model is no smaller than $-2L$ when $\langle J_{ij}^2\rangle=1/L$.
Thus, we set the shift as $2L$ to guarantee that $f > 0$. 

We still need to consider the effect of the
connection matrix.
We use extreme value theory.  We consider
that the matrix elements of $J$ are randomly chosen, so that $H$ becomes sum of $LC$ Gaussian-like variables instead of $L^2$ variables, so the minimum is $\sqrt{C/L}$ times that of the minimum without the connection matrix. To normalize the minimum back to $-2L$, we finally choose $\langle J_{ij}^2\rangle=1/C$. 

We calculate  $\langle H^2\rangle$ for the planar Potts model. When $q=2$, 
the average value 
of $\cos^2\theta_{ij}$ is $1$, while when $q>2$, the value is $1/2$. 
Thus,
\begin{eqnarray}
\nonumber\langle H^2\rangle&=&\left<\sum_{ij}J_{ij}\Delta_{ij}\cos\theta_{ij}\sum_{kl}J_{kl}\Delta_{kl}\cos\theta_{kl}\right>
\\\nonumber&=&2\left<\sum_{ij}J_{ij}^2\Delta_{kl}\cos^2\theta_{ij}\right>
\\\nonumber&=&\sum_{ij}\langle J_{ij}^2\rangle(1+\delta_{q,2})\Delta_{kl}
\\&=&L(1+\delta_{q,2})
\end{eqnarray}

\subsection{The GNK Model}
We use extreme value theory to discuss the minimum and normalization of the GNK model. When $\langle\sigma^2\rangle=2 p!/C$, 
\begin{equation}
H\sim p!N(0,\langle \sigma^2 \rangle LC/p!)\sim N(0,2L)
\end{equation}
so the distribution of the $H$ for the GNK model is the same as that of the $p$SK model.
The correlation induced by the dependence of $H$ on the $s_i$ is, however,
different in these two models.
 Here we follow the method of \cite{pisier1999volume}. We have two sets of variables with the same set size, one is the $p$ spin interaction GNK $H$, denoted as $X_i$, and normalized so that $\langle X_i^2\rangle=1$, and the other is a set of Gaussian  variables $Y_i$, and $\langle Y_i^2\rangle=1$, $\langle Y_iY_j\rangle=c<1$. There are $q^L$ variables in each set, and we know that $\langle Y_{\text{max}}\rangle$ is $\sqrt{1-c}$ times the maximum of the same number of uncorrelated Gaussian variables with the same variance and average \cite{owen1962moments}, so from Eq.\ (\ref{eq:min}) $\langle Y_{\text{max}}\rangle=\sqrt{2L(1-c)\ln q}$. 
We define a new variable, $Z_i=\sqrt{1-t}X_i+\sqrt{t}Y_i$, where $0\le t\le 1$, and we define $r(t)=P(Z_1(t)\le a,...,Z_{q^L}(t)\le a)$. So $r(t=0)$ is the probability that the maximum of  $X_i$ is smaller than $a$, while $r(t=1)$ is the probability that the maximum of  $Y_i$ is smaller than $a$. We seek to make $r(0)\le r(1)$, so that $X_{\text{max}}\ge Y_{\text{max}}$, so we seek $dr(t)/dt\ge0$ for $0\le t\le1$. It is shown that on page 71 of  \cite{pisier1999volume}:
\begin{equation}\label{eq:dp}dr(t)/dt=\frac{1}{2}\sum_{ij}(\langle Y_iY_j\rangle-\langle X_iX_j\rangle)\int^{a}_{-\infty}\cdot\cdot\cdot\int^{a}_{-\infty}\frac{\partial^2\phi(t,Z_1,...,Z_{q^L})}{\partial Z_i\partial Z_j}dZ_1...dZ_{q^L}\end{equation}
where $\phi$ is the joint distribution of $Z$. For example, the term corresponding to $i=1$, $j=2$ is $(\langle Y_1Y_2\rangle-\langle X_1X_2\rangle)\int^{a}_{-\infty}\cdot\cdot\cdot\int^{a}_{-\infty}\phi(t,a,a,Z_3,...,Z_{q^L})dZ_3...dZ_{q^L}$, and $\int^{a}_{-\infty}\cdot\cdot\cdot\int^{a}_{-\infty}\phi(t,a,a,Z_3,...,Z_{q^L})dZ_3...dZ_{q^L}\approx \phi(t,Z_1=a,Z_2=a)$, as the probability that all other variables is smaller than the maximum we are looking for is approximately $1$. $\langle Y_iY_j\rangle=c$ for all $i\ne j$, and $\langle X_iX_j\rangle$ is the same for pairs with $d_{ij}$ the same, so we group the pairs according to their Hamming distance, and rewrite Eq.\ (\ref{eq:dp}) as
\begin{equation}
dr(t)/dt=\frac{1}{4}\sum_{i}\sum_{d=1}^{L}\binom{L}{d}(q-1)^d(c-\langle X_iX_j\rangle)\phi(t,Z_i=a,Z_j=a)
\end{equation}
where $X_j$ is any sequence satisfying $d_{ij}=d$. As the system is totally random, it does not matter what sequence $j$ is, and we can set it as the first sequence and rewrite it as
\begin{equation}\label{eq:dpdt}
dr(t)/dt=\frac{1}{4}q^L\sum_{d=1}^{L}\binom{L}{d}(q-1)^d(c-\langle X_1X_i\rangle)\phi(t,Z_1=a,Z_i=a)
\end{equation}
where $X_i$ is any sequence satisfying $d_{1i}=d$. As only the integral depends on $t$, we can write it as
\begin{equation}\label{eq:int}
r(1)-r(0)=\frac{1}{4}q^L\sum_{d=1}^{L}\binom{L}{d}(q-1)^d(c-\langle X_1X_i\rangle)\int_{0}^1\phi(t,Z_1=a,Z_i=a)
\end{equation}
when $r(0)=r(1)$, we find
\begin{equation}\label{eq:semi}
c=\frac{\sum_{d=1}^{L}\binom{L}{d}(q-1)^d\int_{0}^1\phi(t,Z_1=a,Z_i=a)dt\langle X_1X_i\rangle}{\sum_{d=1}^{L}\binom{L}{d}(q-1)^d\int_{0}^1\phi(t,Z_1=a,Z_i=a)dt}
\end{equation}

We first calculate $\langle X_iX_j\rangle$. The correlation is 
\begin{eqnarray}
\langle X_iX_j\rangle&=&\sum_{i_1...i_p}\sigma({s_{i_1}...s_{i_p}})\Delta_{i_1...i_p}\sum_{j_1...j_p}\sigma({s'_{j_1}...s'_{j_p}})\Delta_{j_1...j_p}\nonumber
\\&=&(p!)^2\sum_{i_1<...<i_p}\sigma({s_{i_1}...s_{i_p}})\sigma({s'_{i_1}...s'_{i_p}})\Delta_{i_1...i_p}
\end{eqnarray}
We initially neglect the $\Delta$, as among the $\binom{L}{p}$ different kinds of $\sigma$, there are $\binom{L-d}{p}$ remaining unchanged, $\langle X_iX_j\rangle=\langle X_i^2\rangle\binom{L-d}{p}/\binom{L}{p}=\binom{L-d}{p}/\binom{L}{p}$ if $d\le L-p$, otherwise it is $0$. Now considering the connection matrix, if $M=0$, the connections are randomly chosen, so we expect that the result does not change. If $M=1$, all connections fall into small modules, and again the $d$ changed spins are randomly distributed in each module, so out of the $l$ spins in each module, $dl/L$ spins have changed. Thus, in each module, out of the $\binom{l}{p}$ spins, $\binom{l-dl/L}{p}$ remain unchanged, so $\langle X_iX_j\rangle=\langle X_i^2\rangle\binom{l-dl/L}{p}/\binom{l}{p}\approx (l-dl/L)^p/l^p\approx (L-d)^p/L^p\approx \binom{L-d}{p}/\binom{L}{p}$. For general $M$, for a given $\sigma$, consider that $h$ spins fall out of the module, and $p-h$ will fall in the module. Among the $\binom{L-l}{h}$ possibilities outside of the module, $\binom{(L-l)(L-d)/L}{h}$ remain unchanged, so the ratio of unchanged over all is is $\binom{(L-l)(L-d)/L}{h}/\binom{L-l}{h}\approx (L-d)^h/L^h$. For the $\binom{l}{p-h}$ possible choices inside the module, $\binom{l(L-d)/L}{p-h}$ is unchanged, so the ratio is $\binom{l(L-d)/L}{p-h}/\binom{l}{p-h}\approx (L-d)^{p-h}/L^{p-h}$. So the probability that a $\sigma$ is unchanged is $(L-d)^h/L^h\times(L-d)^{p-h}/L^{p-h}=(L-d)^p/L^p\approx\binom{L-d}{p}/\binom{L}{p}$. So for all $M$, $\langle X_iX_j\rangle=\langle X_i^2\rangle\binom{L-d}{p}/\binom{L}{p}=\binom{L-d}{p}/\binom{L}{p}$. Also, $\binom{L}{d}\binom{L-d}{p}/\binom{L}{p}=\binom{L-p}{d}$, so
\begin{equation}\label{eq:final}
c=\frac{\sum_{d=1}^{L-p}\binom{L-p}{d}(q-1)^d\int_{0}^1\phi(t,Z_1=a,Z_i=a)dt}{\sum_{d=1}^{L}\binom{L}{d}(q-1)^d\int_{0}^1\phi(t,Z_1=a,Z_i=a)dt}
\end{equation}
The integral part simplifies as
  $$\int^{a}_{-\infty}\cdot\cdot\cdot\int^{a}_{-\infty}\phi(t,a,Z_2,...,Z_{i-1},a,Z_{i+1},...,Z_{q^L})dZ_2...dZ_{i-1}dZ_{i+1}...dZ_{q^L}\approx\phi(t,Z_1=a,Z_i=a)$$
We write the joint distribution of the Gaussian variables as
\begin{equation}\label{eq:multi}
f(x_1,...,x_k)=\frac{1}{\sqrt{(2\pi)^k}\vert \Sigma \vert}e^{-\frac{1}{2}(\boldsymbol{Z-\langle Z \rangle })^{T}\Sigma^{-1}(\boldsymbol{Z-\langle Z \rangle})}
\end{equation}
where $\Sigma$ is the covariance matrix, and $ \vert \Sigma \vert$ is the determinant of $\Sigma$. For our particular case, where there are only two variables with unit variance and zero average, the covariance matrix is given by
\begin{equation}
\Sigma=\begin{pmatrix}
  \langle Z_1^2 \rangle & \langle Z_1 Z_2 \rangle) \\
  \langle Z_1 Z_2 \rangle& \langle Z_2^2  \rangle 
 \end{pmatrix}
 =\begin{pmatrix*}
  1 & \rho \\
  \rho& 1 
 \end{pmatrix*}
\end{equation}
where $\rho= \langle Z_1 Z_2 \rangle $. So $\vert \Sigma \vert=1-\rho^2$ and
\begin{equation}
\Sigma^{-1}=\frac{1}{1-\rho^2}\begin{pmatrix}
  1 & -\rho \\
  -\rho&1 
 \end{pmatrix}
\end{equation}
It follows that
\begin{equation}
-\frac{1}{2}(\boldsymbol{Z-\langle Z \rangle})^{T}\Sigma^{-1}(\boldsymbol{Z-\langle Z \rangle})=-\frac{Z_1^2+Z_2^2-2\rho Z_1 Z_2}{2-2\rho^2}
\end{equation}
When both variables equal $a$, it is $-\frac{a^2}{1+\rho}$. So putting everything into Eq.\ (\ref{eq:multi}), the probability that both variables are $a$ is $e^{-a^2/(1+\rho)}/\sqrt{1-\rho^2}\pi$. We calculate $\rho_{ij}(t)=[(1-t)\langle X_iX_j\rangle+t\langle Y_iY_j\rangle]/(1-t+t)=tc+(1-t)\binom{L-d}{p}/\binom{L}{p}=tc+(1-t)(1-\frac{d}{L})^p$. 

Note that $a=\sqrt{2L(1-c)\ln q}$, so Eq.\ (\ref{eq:final}) is a self-consistent equation
\begin{equation}\label{eq:ff}
c=\frac{\sum_{d=1}^{L-p}\binom{L-p}{d}(q-1)^d\int_{0}^1e^{-2L(1-c)\ln q/[1+tc+(1-t)(1-\frac{d}{L})^p]}/\sqrt{1-[tc+(1-t)(1-\frac{d}{L})^p]^2}dt}{\sum_{d=1}^{L}\binom{L}{d}(q-1)^d\int_{0}^1e^{-2L(1-c)\ln q/[1+tc+(1-t)(1-\frac{d}{L})^p]}/\sqrt{1-[tc+(1-t)(1-\frac{d}{L})^p]^2}dt}
\end{equation}
The solution for $L=100$,  $p=2$, and $q=2$ is $c=0.331$, so $\sqrt{1-c}=0.82$. With $p=2$, $q=2$, and $L=1000$, $c=0.330$, the value of $c$ near that for infinite $L$. We also applied this method to the $p$SK model. For $p=2$, $q=2$, the correlation is $1-4d(L-d)/L^2$. So
\begin{equation}
c=\frac{\sum_{d=1}^{L}[1-4d(L-d)/L^2]\int_{0}^1e^{-2L(1-c)\ln q/[1+tc+(1-t)(1-\frac{d}{L})^p]}/\sqrt{1-[tc+(1-t)(1-\frac{d}{L})^p]^2}dt}{\sum_{d=1}^{L}\binom{L}{d}\int_{0}^1e^{-2L(1-c)\ln q/[1+tc+(1-t)(1-\frac{d}{L})^p]}/\sqrt{1-[tc+(1-t)(1-\frac{d}{L})^p]^2}dt}
\end{equation}
When $p=2$, $q=2$, and $L=100$ we find $c=0.128$, and $H_{\text{min}}=0.777\times 2L$, quite near the numerical result $H_{\text{min}}=0.763\times 2L$.
This self-consistent method, thus, is fairly accurate.

For a given $q$, larger $p$ lead to smaller $c$ using Slepian's lemma Eq.\ (\ref{eq:slepian}). To prove this, first  assume $V_i=H^{\text{GNK}}_{p_v,q}(S_i)$ and $W_i=H^{\text{GNK}}_{p_w,q}(S_i)$ are GNK model variables with $p_V>p_W$. For any sequence $S_i$, we group all other sequences according to the Hamming distance between them, so the group with Hamming distance $d$ contains $\binom{L}{d}(q-1)^{d}$ variables. Assume $S_j$ has a Hamming distance $d$ from $S_i$, so $\langle V_iV_j\rangle=\binom{L-d}{p_V}/\binom{L}{p_V}=\binom{L-p_V}{d}/\binom{L}{d}$, $\langle W_iW_j\rangle=\binom{L-d}{p_W}/\binom{L}{p_W}=\binom{L-p_W}{d}/\binom{L}{d}$. As  $p_V>p_W$, $\binom{L-p_V}{d}<\binom{L-p_W}{d}$, so $\langle V_iV_j\rangle<\langle W_iW_j\rangle$. As the $S_i$  is chosen randomly, the correlation between any pair of $V$ is smaller than that of the corresponding pair of $W$, and according to Slepian's lemma, the minimum of $V$ is smaller. Thus, it is proved that larger $p$ lead to smaller minima.

We calculate the results for $p=2$ of different $q$. Shown in Table \ref{table:re} is the result of $L=100$, $p=2$ and $3\le q\le 20$. We see that $c$ decreases monotonically for increasing $q$ in this range.  
\begin{table}[tb!]
\caption{
 \label{table:re}
Self-consistent calculation of $c$ as a function of $q$ using
Eq.\ (\ref{eq:ff}). Here $L=100$ and $p=2$.}
\begin{tiny}
\begin{tabular}{l|llllllllllllllllll}
 $q$  &3  &4  &5&6&7&8&9&10&11  &12  &13  &14  &15&16&17&18&19&20 \\
\hline\\
 $c$& 0.219 & 0.17 & 0.15 & 0.129&0.116  & 0.107 & 0.100 & 0.094 & 0.089 
 & 0.085 & 0.082 &0.079  &0.076 & 0.074 &0.072  & 0.070 &  0.068& 0.067\\
\hline
\end{tabular}
\end{tiny}
\end{table}
We also calculated $c$ for large $q$ and $L=1000000$, Table \ref{table:re1}.
\begin{table}[tb!]
\caption{Self-consistent calculation of $c$ as function of $q$  using 
Eq.\ (\ref{eq:ff}). Here $L=1000000$ and $p=2$.\label{table:re1}}
\begin{tabular}{l|lllllllll}
 $q$  &40  &80  &160&320&640&1280&2560&5120&10240 \\
\hline\\
 $c$& 0.049 & 0.038 & 0.031 & 0.026&0.022  & 0.019& 0.016& 0.014 &0.013
\end{tabular}
\end{table}

It is apparent that all results for $c$ fall below $0.33$, so the minimum of GNK model will fall between $0.82\times\sqrt{2L\ln q}$ and $\sqrt{2L\ln q}$ for any $p$ and $q$. As $0.82\approx 1$, we neglect this factor, i.e.\ conservatively assure that the fitness is positive, not merely non-negative. When $\langle H^2\rangle=2L/\ln q$, the minimum will be $2L$.

In summary,
for all models discussed in this appendix, a shift of $2L$ 
assures that the fitness is positive.

\section{Appendix B: Calculation of Taylor Series Expansion for Average Fitness}\label{sec:details}
We here describe how the coefficients in the Taylor series
expansion of $\langle f(t) \rangle $ are calculated.

\subsection{Organizing the Terms in the Taylor Series Expansion}\label{sec:divtay}
We define $f(\{n_a\})=\sum_{n_i}f(S_i)n_i/N$, the average fitness of an individual of configuration $\{n_a\}$. The first order result of Eq.\ (\ref{2}) is a linear combination of $P(\{n_a\},t)$, which can be divided into three parts: $f$ part, from natural selection; $\mu$  part, from mutation; $\nu$ part, from horizontal gene transfer. In a compact form
\begin{equation}
\frac{dP}{dt}=\mathcal{L}P=\mathcal{L}_fP+\mathcal{L}_\mu P+\mathcal{L}_\nu P
\end{equation}
and
\begin{equation}
\frac{d\langle f\rangle}{dt}=\sum_{\{n_a\}}f(\{n_a\})\mathcal{L}P(\{n_a\},t)
\end{equation}
In this way we obtain the form of higher order results, for example, the second order result can be expressed as
\begin{equation}
\frac{d^2\langle f\rangle}{dt^2}=\frac{d}{dt}\frac{d\langle f\rangle}{dt}=\sum_{\{n_a\}}f(\{n_a\})\mathcal{L}^2P=\sum_{\{n_a\}}f(\{n_a\})(\mathcal{L}_f+\mathcal{L}_\mu+\mathcal{L}_\nu)(\mathcal{L}_f+\mathcal{L}_\mu+\mathcal{L}_\nu)P
\end{equation}
which is divided into $9$ terms:
\begin{eqnarray}
\nonumber\frac{d^2\langle f\rangle}{dt^2}&=&\sum_{\{n_a\}}f(\{n_a\})(\mathcal{L}_f\mathcal{L}_f+\mathcal{L}_f\mathcal{L}_\mu+\mathcal{L}_f\mathcal{L}_\nu+\mathcal{L}_\mu \mathcal{L}_f
\\&&+\mathcal{L}_\mu\mathcal{L}_\mu+\mathcal{L}_\mu\mathcal{L}_\nu+\mathcal{L}_\nu\mathcal{L}_f+\mathcal{L}_\nu\mathcal{L}_\mu+\mathcal{L}_\nu\mathcal{L}_\nu)P
\end{eqnarray}
Note that $\mathcal{L}_f$, $\mathcal{L}_\mu$, $\mathcal{L}_\nu$ are not commutative with each other (but of course commutative with itself), for example $\mathcal{L}_f\mathcal{L}_\mu\neq\mathcal{L}_\mu \mathcal{L}_f$, since these two correspond to different evolutionary processes of the population. As $2b=\frac{d^2\langle f(t)\rangle}{dt^2}|_{t=0}$, we can divide $2b$ into nine terms, and name them $T_{ff}$, $T_{f\mu}$, $T_{f\nu}$, $T_{\mu f}$, $T_{\mu\mu}$, $T_{\mu\nu}$, $T_{\nu f}$, $T_{\nu \mu}$, $T_{\nu\nu}$. Also note that the naming is in a sense inverted, for example $T_{f\mu}$ term, which comes from $\mathcal{L}_f\mathcal{L}_\mu P$, actually represents the process that a mutation happens before natural selection.

In this way, the third order term consists of $27$ terms, and the $r$-th order term consists $3^r$ terms in general. Each term can be named according to the order of $\mathcal{L}$ operators involved. Below we will discuss how to compute each part of an $r$-th order term. First we discuss eliminating terms that do not contribute, then we calculate terms for different models. We will start with the simplest $2$-spin, $2$-state model, and then discuss $p$SK model, planar Potts model, and GNK model.

\subsection{Identification of the Terms that Vanish}
\label{sec:elim}
The number of terms increases exponentially with order, so we would like to eliminate some terms that vanish from some considerations. We discuss this according to the initial conditions, that is, random initial sequences and identical initial sequences.

For both initial conditions, a term that does not contain natural selection processes vanishes. The reason is that the Taylor expansion results correspond to \textit{change} of fitness, so it has a factor of $\Delta f$ proportional to some linear combinations of $H$, which is in turn some linear combinations of $J$ or $\sigma$. For terms that only contain mutations and horizontal gene transfers, the result is proportional to linear combinations of $J$ or $\sigma$, which vanish when we take an average. Only when there is natural selection can $J$ or $\sigma$ be multiplied into second order or higher even order forms, which are non-zero when averaged. So only terms containing natural selections can contribute.

For the random initial sequences, the modular term appears in second order. From the 
principle above, the first order terms $T_{\mu}$ and $T_{\nu}$, and the second order 
terms $T_{\mu\mu}$, $T_{\mu\nu}$, $T_{\nu\mu}$ and $T_{\nu\nu}$ are all $0$. In addition, for the $T_{f\mu}$ term, first an individual mutates then a natural selection happens. As the initial system is totally random, the mutation in the first step does not change the system, so this term is $0$. Then, up to second order, there are five non-zero terms: $T_f$, $T_{ff}$, $T_{f\nu}$, $T_{\mu f}$ and $T_{\nu f}$.

For the same initial sequences, the modular terms appear in the fourth order, meaning that we need to calculate $3+9+27+81=120$  terms. Fortunately, the initial conditions are so special that we can greatly reduce our burden. As discussed above, terms not containing $f$ are automatically zero, so  $120-2-4-8-16=90$  terms can be non-zero. In addition, if a natural selection or horizontal gene transfer process happens first without a mutation, then nothing is changed as every sequence is initially the same, making this term zero. So any term not ending with $\mu$ is zero, leaving behind $0+1+5+19=25$ non-zero terms: $0$ first order term, $1$ second order term $T_{f\mu}$, $5$ third order terms satisfying three-letter-array ending with $\mu$ and containing a $f$ in the first two letters and $19$ fourth order terms satisfying four-letter-array ending with $\mu$ and containing a $f$ in the first three letters. Additionally, for a term starting with $\nu$ (the last process is horizontal gene transfer), if there were fewer than 2 mutational processes before horizontal gene transfer, then the term is zero. The reason is that if only one mutation happened, all mutated individuals are the same, so a horizontal gene transfer changes the population only when it involves a mutated individual and an unmutated individual. And for a change to occur, the horizontal gene transfer must transfer the part of the sequence that is mutated. The probability of the unmutated one becoming a mutated one through horizontal gene transfer is the same as the mutated one becoming an unmutated one, and the change of fitness of these two processes cancel, so the contribution is zero. In this way, $T_{\nu f\mu}$ is zero, and $T_{\nu ff\mu}$, $T_{\nu f \nu \mu}$ and $T_{\nu\nu f\mu}$ is zero. An additional two terms, $T_{\mu\nu f\mu}$ and $T_{\nu\mu f\mu}$ are found to be zero after calculation, so there is one non-zero second order term, four non-zero third order terms, and fourteen non-zero fourth order terms, totaling nineteen non-zero terms.

\subsection{The $p=2$ SK Model with Random Initial Sequences}
To illustrate the calculation for
the random initial sequences, we use $T_f$ term as an example.
\begin{eqnarray}\label{eq:51}
\mathcal{L}_f|n_1,...,n_{2^L}\rangle&=&\sum_{ab}\frac{n_an_b}{N}f_a(|n_a+1,...,n_b-1\rangle-|n_a,...,n_b\rangle)
\end{eqnarray}
So
\begin{eqnarray}\label{eq:randmethod}
\nonumber T_f&=&\sum_{\{n_a\}}f(\{n_a\})\mathcal{L}_fP(\{n_a\},0)
\\\nonumber&=&\sum_{\{n_a\}}P(\{n_a\},0)\sum_{ab}\frac{n_an_b}{N}f_a\left[f(|n_a+1,...,n_b-1\rangle)-f(|n_a,...,n_b\rangle)\right]
\\\nonumber&=&\sum_{\{n_a\}}P(\{n_a\},0)\sum_{ab}\frac{n_an_b}{N}f_a\frac{f_a-f_b}{N}
\\\nonumber&=&\left<\sum_{ab}\frac{n_an_b}{N}f_a\frac{f_a-f_b}{N}\right>
\\\nonumber&=&\left<\sum_{ab}\frac{n_an_b}{N}f_a\frac{H_a-H_b}{N}\right>
\\&=&\left<\sum_{ab}\frac{n_an_b}{N}H_a\frac{H_a-H_b}{N}\right>
\end{eqnarray}
The last equation holds because average over first order terms of $H$ is always zero. As the initial sequences are totally random, when we pick out a particular individual $a$, we expect that from the view of $a$, the other $N-1$ individuals are totally random, so their $H$ is uncorrelated with that of $a$. As $\langle H\rangle=0$,  $\sum_bn_b H_aH_b=0$.

So,
\begin{eqnarray}
\nonumber T_f&=&\left<\sum_{ab}\frac{n_an_b}{N}H_a\frac{H_a-H_b}{N}\right>
\\\nonumber &=&\left<\sum_{ab}\frac{n_an_bH_a^2}{N^2}\right>
\\\nonumber&=&\frac{N(N-1)}{N^2}\langle H_a^2 \rangle
\\\nonumber&=&(1-1/N)\sum_{ijkl}\langle J_{ij}J_{kl}\rangle\Delta_{ij}\Delta_{kl}s_is_js_ks_l
\\\nonumber&=&2(1-1/N)\sum_{ij}\Delta_{ij}/C
\\&=&2L(1-1/N)
\end{eqnarray}
and,
\begin{equation}
a=T_f
\end{equation}
Similarly, we can obtain,
\begin{eqnarray}
\nonumber T_{ff}&=&-8L^2(1-1/N)/N
\\\nonumber T_{\mu f}&=&-8\mu L(1-1/N)
\\\nonumber T_{f\nu}&=&-4\nu L(1-1/N)\left[M+(1-M)/K\right]/N
\\ T_{\nu f}&=&4\nu L(1-1/N)(1-3/N)\left(1-1/K\right)(M-1)
\end{eqnarray}
and
\begin{equation}
2b=T_{ff}+T_{\mu f}+T_{f\nu}+T_{\nu f}
\end{equation}

\subsection{The $p=2$ SK model with Identical Initial Sequences}
To illustrate how to calculate the terms for identical initial sequences
we use $T_{f\mu}$ term as an example.
 We assume the initial sequences are $S_0$, and the initial state is $\{n_a\}_0=(N,0,...,0)=|N\delta_{\boldsymbol{e},0}\rangle$. $\mathcal{L}_\mu$ takes the state to
\begin{equation}
\mathcal{L}_\mu|N\delta_{\boldsymbol{e},0}\rangle=\mu N\sum_{\boldsymbol{a}=\partial 0}\left[|(N-1)\delta_{\boldsymbol{e},0}+\delta_{\boldsymbol{e},\boldsymbol{a}}\rangle-|N\delta_{\boldsymbol{e},0}\rangle\right]
\end{equation}
A subsequent $\mathcal{L}_f$ takes the state to
\begin{eqnarray}
\nonumber\mathcal{L}_f\mathcal{L}_\mu|N\delta_{\boldsymbol{e},0}\rangle
&=&\mu N\mathcal{L}_f\sum_{\boldsymbol{a}=\partial 0}\left[|(N-1)\delta_{\boldsymbol{e},0}+\delta_{\boldsymbol{e},\boldsymbol{a}}\rangle-|N\delta_{\boldsymbol{e},0}\rangle\right]
\\&=&\nonumber\mu N\sum_{\boldsymbol{a}=\partial 0}\Bigg[\frac{N-1}{N}f(S_0)(|N\delta_{\boldsymbol{e},0}\rangle-|(N-1)\delta_{\boldsymbol{e},0}+\delta_{\boldsymbol{e},\boldsymbol{a}}\rangle)
\\&&\nonumber+\frac{N-1}{N}f(S_{\boldsymbol{a}})(|(N-2)\delta_{\boldsymbol{e},0}+2\delta_{\boldsymbol{e},\boldsymbol{a}}\rangle-|(N-1)\delta_{\boldsymbol{e},0}+\delta_{\boldsymbol{e},\boldsymbol{a}}\rangle)\Bigg]
\\&=&\nonumber\mu(N-1)\sum_{\boldsymbol{a}=\partial 0}\big[f(S_0)(|N\delta_{\boldsymbol{e},0}\rangle-|(N-1)\delta_{\boldsymbol{e},0}+\delta_{\boldsymbol{e},\boldsymbol{a}}\rangle)
\\&&+f(S_{\boldsymbol{a}})(|(N-2)\delta_{\boldsymbol{e},0}+2\delta_{\boldsymbol{e},\boldsymbol{a}}\rangle-|(N-1)\delta_{\boldsymbol{e},0}+\delta_{\boldsymbol{e},\boldsymbol{a}}\rangle)\big]
\end{eqnarray}
So after two operations, $P(|N\delta_{\boldsymbol{e},0}\rangle,0)=\mu(N-1)\sum_{\boldsymbol{a}=\partial 0}f(S_0)
$, $P(|(N-1)\delta_{\boldsymbol{e},0}+\delta_{\boldsymbol{e},\boldsymbol{a}}\rangle,0)=-\mu(N-1)\sum_{\boldsymbol{a}=\partial 0}(f(S_{\boldsymbol{a}})+f(S_0))$ and $P(|(N-2)\delta_{\boldsymbol{e},0}+2\delta_{\boldsymbol{e},\boldsymbol{a}}\rangle,0)=\mu(N-1)\sum_{\boldsymbol{a}=\partial 0}f(S_{\boldsymbol{a}})$, and 
we find
\begin{eqnarray}\label{eq:fm}
\nonumber T_{f\mu}&=&\sum_{\{n_a\}}f(\{n_a\})\mathcal{L}_f\mathcal{L}_\mu P(\{n_a\},0)
\\&=&\nonumber f(|N\delta_{\boldsymbol{e},0}\rangle)P(|N\delta_{\boldsymbol{e},0}\rangle,0)\\&&\nonumber+f(|(N-1)\delta_{\boldsymbol{e},0}+\delta_{\boldsymbol{e},\boldsymbol{a}}\rangle)P(|(N-1)\delta_{\boldsymbol{e},0}+\delta_{\boldsymbol{e},\boldsymbol{a}}\rangle,0)
\\&&\nonumber+f(|(N-2)\delta_{\boldsymbol{e},0}+2\delta_{\boldsymbol{e},\boldsymbol{a}}\rangle)P(|(N-2)\delta_{\boldsymbol{e},0}+2\delta_{\boldsymbol{e},\boldsymbol{a}}\rangle,0)
\\&=&\nonumber\mu(1-1/N)\sum_{\boldsymbol{a}=\partial 0}[Nf(S_0)^2-(N-1)f(S_0)(f(S_0)+f(S_{\boldsymbol{a}}))
\\&&\nonumber-f(S_{\boldsymbol{a}})(f(S_0)+f(S_{\boldsymbol{a}}))+(N-2)f(S_{\boldsymbol{a}})f(S_{0})+2f(S_{\boldsymbol{a}})^2]
\\&=&\mu(1-1/N)\sum_{\boldsymbol{a}=\partial 0}(f(S_{\boldsymbol{a}})-f(S_0))^2
\end{eqnarray}
in which 
\begin{eqnarray}
\sum_{\boldsymbol{a}=\partial 0}[f(S_{\boldsymbol{a}})-f(S_0)]^2
&=&\nonumber\sum_{\boldsymbol{a}=\partial 0}\sum_{ij}\sum_{kl}J_{ij}J_{kl}\Delta_{ij}\Delta_{kl}[1-(1-2\delta_{ai})(1-2\delta_{aj})]\\&&\times[1-(1-2\delta_{ak})(1-2\delta_{al})]
\end{eqnarray}
if we average over $J_{ij}J_{kl}$,  we find
\begin{eqnarray}
\nonumber\sum_{\boldsymbol{a}=\partial 0}[f(S_{\boldsymbol{a}})-f(S_0)]^2
&=&\nonumber\sum_{\boldsymbol{a}=\partial 0}\sum_{ijkl}\langle J_{ij}J_{kl}\rangle\Delta_{ij}\Delta_{kl}[1-(1-2\delta_{ai})(1-2\delta_{aj})]\\&&\nonumber\times[1-(1-2\delta_{ak})(1-2\delta_{al})]
\\&=&\nonumber\sum_{\boldsymbol{a}=\partial 0}\sum_{ijkl}(\delta_{ik}\delta_{jl}+\delta_{il}\delta_{jk})(1-\delta_{ij})\Delta_{ij}\Delta_{kl}/C
\\&&\nonumber\times[1-(1-2\delta_{ai})(1-2\delta_{aj})][1-(1-2\delta_{ak})(1-2\delta_{al})]
\\&=&\nonumber 2\sum_{\boldsymbol{a}=\partial 0}\sum_{ij}\Delta_{ij}[1-(1-2\delta_{ai})(1-2\delta_{aj})]^2/C
\\&=&\nonumber8\sum_{\boldsymbol{a}=\partial 0}\sum_{ij}\Delta_{ij}(\delta_{ai}+\delta_{aj})/C
\\&=&16L
\end{eqnarray}
So
\begin{equation}
T_{f\mu}=16\mu L(1-1/N)
\end{equation}
and
\begin{equation}
2b=T_{f\mu}
\end{equation}

Similarly, we find
\begin{eqnarray}
\nonumber T_{f\mu\mu}&=&-128\mu^2L(1-1/N)
\\\nonumber T_{f\nu\mu}&=&-32\mu\nu L(1-1/N)/N
\\\nonumber T_{ff\mu}&=&-64\mu L^2(1-1/N)/N
\\\nonumber T_{\mu f\mu}&=&-64\mu^2 L(1-1/N)
\\\nonumber T_{fff\mu}&=&256\mu L^3(1-1/N)/N^2+128\mu L^2(1-1/N)/N^2
\\\nonumber &&+256\mu L(1-1/N)\left(3-\frac{18}{N}+\frac{23}{N^2}\right)
\\\nonumber T_{ff\mu\mu}&=&512\mu^2L^2(1-1/N)/N
\\\nonumber T_{ff\nu\mu}&=&128\mu\nu L^2(1-1/N)/N^2
\\\nonumber T_{f\mu f\mu}&=&512\mu^2L^2(1-1/N)/N
\\\nonumber T_{f\mu\mu\mu}&=&1024\mu^3L(1-1/N)
\\\nonumber T_{f\mu\nu\mu}&=&256\mu^2\nu L(1-1/N)/N
\\\nonumber T_{f\nu f\mu}&=&128\mu\nu L^2(1-1/N)/N^2
\\\nonumber T_{f\nu\mu\mu}&=&128\mu^2\nu L(1-1/N)\left[2+(1-M)\left(1-1/K\right)\right]/N
\\\nonumber T_{f\nu\nu\mu}&=&64\mu\nu^2L(1-1/N)/N^2
\\\nonumber T_{\mu ff\mu}&=&256\mu^2 L^2(1-1/N)/N
\\\nonumber T_{\mu f\mu\mu}&=&512\mu^3 L(1-1/N)
\\\nonumber T_{\mu f\nu\mu}&=&128\mu^2\nu L(1-1/N)/N
\\\nonumber T_{\mu\mu f\mu}&=&256\mu^3 L(1-1/N)
\\ T_{\nu f\mu\mu}&=&-128\mu^2\nu L(1-1/N)(1-3/N)(1-M)\left(1-1/K\right)
\end{eqnarray}
and
\begin{eqnarray}
(3!) c&=&T_{f\mu\mu}+T_{f\nu\mu}+T_{ff\mu}+T_{\mu f\mu}\nonumber
\\\nonumber (4!) d&=&T_{fff\mu}+T_{ff\mu\mu}+T_{ff\nu\mu}+T_{f\mu f\mu}+T_{f\mu\mu\mu}+T_{f\mu\nu\mu}+T_{f\nu f\mu}
\\ &&+T_{f\nu\mu\mu}+T_{f\nu\nu\mu}+T_{\mu ff\mu}+T_{\mu f\mu\mu}+T_{\mu f\nu\mu}+T_{\mu\mu f\mu}+T_{\nu f\mu\mu}
\end{eqnarray}

\subsection{The $p$SK Interaction}
For the $p$-spin interaction, the relationship between different $J_{i_1,...,i_p}$ is quite similar to the $p=2$ case, and the principles eliminating zero terms are the same. Moreover, the physical processes corresponding to each Taylor expansion term remain the same, so we just need to change the form of the $H$. For example, to obtain $T_{f\mu}$ term of same initial conditions, Eq.\ (\ref{eq:fm}) still holds, but we need to use the new $H$ to calculate $\sum_{\boldsymbol{a}=\partial 0}[f(S_{\boldsymbol{a}})-f(S_0)]^2$. For the new form,
\begin{eqnarray}
\nonumber\sum_{\boldsymbol{a}=\partial 0}[f(S_{\boldsymbol{a}})-f(S_0)]^2
&=&\nonumber\sum_{\boldsymbol{a}=\partial 0}\sum_{i_1...i_pj_1...j_p}\langle J_{i_1...i_p}J_{j_1...j_p}\rangle\Delta_{i_1...i_p}\Delta_{j_1...j_p}\\\nonumber &&\times[1-\Pi_{q=1}^p(1-2\delta_{a,i_q})][1-\Pi_{q=1}^p(1-2\delta_{a,j_q})]
\\&=&\nonumber\sum_{\boldsymbol{a}=\partial 0}\sum_{i_1...i_p}p!\frac{2}{Cp!}\left(2\sum_{q=1}^p\delta_{a,i_q}\right)^2\Delta_{i_1...i_p}
\\&=&\nonumber \frac{8}{C}\sum_{\boldsymbol{a}=\partial 0}\sum_{i_1...i_p}\Delta_{i_1...i_p}\sum_{i=1}^p\delta_{a,i_q}
\\&=&\nonumber \frac{8p}{C}\sum_{i_1...i_p}\Delta_{i_1...i_p}
\\&=&8pL
\end{eqnarray}
So $T_{f\mu}=8\mu pL(1-1/N)$. In this way, we can obtain for $p$-spin interaction random initial sequences,
\begin{eqnarray}
\nonumber T_f&=&2L(1-1/N)
\\\nonumber T_{ff}&=&-8L^2(1-1/N)/N
\\\nonumber T_{\mu f}&=&-4\mu pL(1-1/N)
\\\nonumber T_{f\nu}&=&-4\nu L(1-1/N)\left(M+\left(\frac{1}{K}\right)^{p-1}(1-M)\right)/N
\\ T_{\nu f}&=&2p\nu L(1-1/N)(1-3/N)\left(1-\left(\frac{1}{K}\right)^{p-1}\right)(M-1)
\end{eqnarray}
and
\begin{eqnarray}
a&=&T_f\nonumber
\\2b&=&T_{ff}+T_{\mu f}+T_{f\nu}+T_{\nu f}
\end{eqnarray}
Also, we find for initial conditions of identical sequences,
\begin{eqnarray}
\nonumber T_{f\mu\mu}&=&-64\mu^2 pL(1-1/N)
\\\nonumber T_{f\nu\mu}&=&-16\mu\nu pL(1-1/N)/N
\\\nonumber T_{ff\mu}&=&-32\mu pL^2(1-1/N)/N
\\\nonumber T_{\mu f\mu}&=&-16\mu^2 p^2L(1-1/N)
\\\nonumber T_{fff\mu}&=&128\mu pL^3(1-1/N)/N^2+64\mu pL^2(1-1/N)/N^2
\\\nonumber &&64\mu p^2L(1-1/N)\left(3-\frac{18}{N}+\frac{23}{N^2}\right)
\\\nonumber T_{ff\mu\mu}&=&128\mu^2p^2L^2(1-1/N)/N
\\\nonumber T_{ff\nu\mu}&=&64\mu\nu pL^2(1-1/N)/N^2
\\\nonumber T_{f\mu f\mu}&=&128\mu^2p^2L^2(1-1/N)/N
\\\nonumber T_{f\mu\mu\mu}&=&128\mu^3p^3L(1-1/N)
\\\nonumber T_{f\mu\nu\mu}&=&64\mu^2\nu p^2L(1-1/N)/N
\\\nonumber T_{f\nu f\mu}&=&64\mu\nu pL^2(1-1/N)/N^2
\\\nonumber T_{f\nu\mu\mu}&=&64\mu^2\nu L(1-1/N)\left[p^2+p(p-1)(1-M)\left(1-1/K\right)\right]/N
\\\nonumber T_{f\nu\nu\mu}&=&32\mu\nu^2pL(1-1/N)/N^2
\\\nonumber T_{\mu ff\mu}&=&64\mu^2 p^2L^2(1-1/N)/N
\\\nonumber T_{\mu f\mu\mu}&=&64\mu^3 p^3L(1-1/N)
\\\nonumber T_{\mu f\nu\mu}&=&32\mu^2\nu p^2L(1-1/N)/N
\\\nonumber T_{\mu\mu f\mu}&=&32\mu^3 p^3L(1-1/N)
\\ T_{\nu f\mu\mu}&=&-64\mu^2\nu p(p-1)L(1-1/N)(1-3/N)(1-M)\left(1-1/K\right)
\end{eqnarray}
and
\begin{eqnarray}
a&=&T_f\nonumber
\\(2!) b&=&T_{ff}+T_{\mu f}+T_{f\nu}T_{\nu f}\nonumber
\\(3!) c&=&T_{f\mu\mu}+T_{f\nu\mu}+T_{ff\mu}+T_{\mu f\mu}\nonumber
\\\nonumber (4!) d&=&T_{fff\mu}+T_{ff\mu\mu}+T_{ff\nu\mu}+T_{f\mu f\mu}+T_{f\mu\mu\mu}+T_{f\mu\nu\mu}+T_{f\nu f\mu}
\\&&+T_{f\nu\mu\mu}+T_{f\nu\nu\mu}+T_{\mu ff\mu}+T_{\mu f\mu\mu}+T_{\mu f\nu\mu}+T_{\mu\mu f\mu}+T_{\nu f\mu\mu}
\end{eqnarray}

\subsection{The Planar Potts Model with Random Initial Sequences}

The non-zero terms are still $T_f$, $T_{ff}$, $T_{\mu f}$, $T_{f\nu}$ and $T_{\nu f}$. As $\langle H^2\rangle$ is changed, all terms will also change. In addition, the $T_{\mu f}$ term has a mutational process, which depends on the number of directions a spin can have. We assume that after a mutation, a spin randomly chooses a \textit{different} direction as before. Using methods similar to Eq.\ (\ref{eq:randmethod}), we find $T_{\mu f}=\mu \sum_{\partial a}(1-1/N)\left< H_a(H_{\partial a}-H_a)\right>$.  Considering the different directions, we find
\begin{eqnarray}
\nonumber&&\sum_{\partial a}\left< H_a(H_{\partial a}-H_a)\right>\\\nonumber&=&\sum_{\partial a}\left<\sum_{ij}J_{ij}\Delta_{ij}s_is_j\cos\theta_{ij}\sum_{kl}J_{kl}\Delta_{kl}s_ks_l(\cos\theta_{kl}'-\cos\theta_{kl})\right>
\\&=&\sum_{\partial a}\left<\sum_{ij}2\frac{1}{C}\Delta_{ij}\cos\theta_{ij}(\cos\theta_{ij}'-\cos\theta_{ij})\right>
\end{eqnarray}
where $\theta_{ij}'$ is the same as $\theta_{ij}$ if neither $i$ nor $j$ is mutated, otherwise $\theta_{ij}'$ has the same probability to be any allowed value other than  $\theta_{ij}$. So we only need to consider the case when either $i$ or $j$ is mutated, and obtain an extra factor of $2$ from this,
\begin{eqnarray}
&&\nonumber\sum_{\partial a}\left<\sum_{ij}2\frac{1}{C}\Delta_{ij}\cos\theta_{ij}(\cos\theta_{ij}'-\cos\theta_{ij})\right>\\&=&\frac{4}{C}\left<\sum_{ij}\Delta_{ij}\cos\theta_{ij}(\cos\theta_{ij}'-\cos\theta_{ij})\right>
\end{eqnarray}
and
\begin{eqnarray}
\nonumber\langle\cos\theta_{ij}\cos\theta_{ij}'\rangle&=&\left<\frac{1}{q-1}\sum_{r\neq s}\cos \theta_r\cos\theta_s\right>
\\\nonumber&=&\left<\frac{1}{q-1}(\sum_{r}\cos \theta_r-\cos\theta_s)\cos\theta_s\right>
\\&=&-\frac{1}{q-1}\langle\cos^2\theta_s\rangle
\end{eqnarray}
Then,
\begin{eqnarray}
\nonumber\left< H_a(H_{\partial a}-H_a)\right>&=&\nonumber\left<\frac{4}{C}\sum_{ij}\Delta_{ij}\cos\theta_{ij}(\cos\theta_{ij}'-\cos\theta_{ij})\right>
\\\nonumber&=&\frac{4}{C}\sum_{ij}\Delta_{ij}(-\frac{1}{q-1}-1)\langle\cos^2\theta\rangle
\\\nonumber&=&-4L\frac{q}{q-1}\langle\cos^2\theta\rangle
\\&=&-2L\frac{q}{q-1}(1+\delta_{q,2})
\end{eqnarray}
So 
\begin{equation}
T_{\mu f}=-2\mu L(1-1/N)\frac{q}{q-1}(1+\delta_{q,2})
\end{equation}
The other terms are
\begin{equation}a=T_f=L(1-1/N)(1+\delta_{q,2})\end{equation}
\begin{eqnarray}
\nonumber T_{ff}&=&-4(1+\delta_{q,2})L^2(1-1/N)/N
\\\nonumber T_{\mu f}&=&-2(1+\delta_{q,2})\mu L(1-1/N)\frac{q}{q-1}
\\\nonumber T_{f\nu}&=&-2(1+\delta_{q,2})\nu L(1-1/N)[M+(1-M)/K]/N
\\ T_{\nu f}&=&2(1+\delta_{q,2})\nu L(1-1/N)(1-3/N)(1-1/K)(M-1)
\end{eqnarray}
and
\begin{equation}
2b=T_{ff}+T_{\mu f}+T_{f\nu}+T_{\nu f}
\end{equation}

\subsection{GNK Model Calculation}

The processes corresponding to different terms are still the same, but for the GNK model, as the $H$ is quite different, the Taylor expansion results will change. For example, for two randomly chosen sequences $S_a$ and $S_b$, their correlation will be non-zero. Instead, it will be
\begin{eqnarray}
\nonumber\langle H_aH_b\rangle&=&\left<\sum_{i_1...i_p}\sigma_{i_1...i_p}(s_{i_1}^a...s_{i_p}^a)\Delta_{i_1...i_p}\sum_{j_1...j_p}\sigma_{j_1...j_p}(s_{i_1}^b...s_{i_p}^b)\Delta_{j_1...j_p}\right>
\\\nonumber&=&(p!)^2\sum_{i_1<...<i_p}\langle\sigma_{i_1...i_p}(s_{i_1}^a...s_{i_p}^a)\sigma_{i_1...i_p}(s_{i_1}^b...s_{i_p}^b)\rangle\Delta_{i_1...i_p}
\\\nonumber&=&(p!)^2\sum_{i_1<...<i_p}\frac{1}{q^p}\langle\sigma^2\rangle\Delta_{i_1...i_p}
\\\nonumber&=&p!\sum_{i_1...i_p}\frac{1}{q^p}\langle\sigma^2\rangle\Delta_{i_1...i_p}
\\\nonumber&=&p!\frac{1}{q^p}\frac{2}{p!C\ln q}LC
\\&=&\frac{2L}{q^p\ln q}
\end{eqnarray}
where the $1/q^p$ factor comes from the fact that only when the states of all corresponding spins match can the correlations of $\sigma$ be nonzero.

Similarly we calculate all terms and obtain
\begin{equation}
a=T_f=2L(1-1/N)(1-1/q^p)/\ln q
\end{equation}
and
\begin{eqnarray}
\nonumber 2b&=&T_{ff}+T_{\mu f}+T_{f\nu}+T_{\nu f}
\\\nonumber T_{ff}&=&-\frac{8L^2(1-1/N)}{N\ln q}(1-1/q^p)
\\\nonumber T_{\mu f}&=&-2\mu pL(1-1/N)/\ln q
\\\nonumber T_{f\nu}&=&\frac{4\nu K(1-1/N)}{N\ln
q}\left[L(1-M)\left(\frac{1-1/K+q/K}{q}\right)^p+LM/K+L(1-1/K)M/q^p\right]
\\\nonumber T_{\nu f}&=&\frac{2\nu K}{\ln q}(M-1)(1-1/N)(1-3/N)
\\&&\times\left[1+1/q^p-\left(\frac{1-1/K+q/K}{q}\right)^p-\left(\frac{q-q/K+1/K}{q}\right)^p\right]
\end{eqnarray}

\section{Appendix C: Calculation of the Taylor Series Expansion
for the Sequence Divergence}\label{sec:detailsdiv}
We here describe how the Taylor expansion series of sequence divergence Eq.\ (\ref{4}) are
calculated.

As the divergence is $D = \frac{1}{N}\sum_{\alpha=1}^N
\langle
\frac{L - S_\alpha(t) \cdot S_\alpha(0)}{2}
\rangle
$, it is determined by the changes to the sequences of the population, which can be
tracked using Eq.\ (\ref{2}). The order of the terms corresponds to the number of
processed involved, for example, second order terms involve two processes. Using conventions 
developed in section \ref{sec:divtay}, we
divide the terms according to what evolutionary processes it involves, for example, the
term which is the result of a mutational process followed by a horizontal gene transfer is
called $D_{\nu\mu}$, similar to the naming norm used in section \ref{sec:divtay}. 
So
\begin{eqnarray}
\alpha&=&D_f+D_{\mu}+D_{\nu}\nonumber
\\(2!)\beta&=&D_{ff}+D_{f\mu}+D_{f\nu}+D_{\mu
f}+D_{\mu\mu}+D_{\mu\nu}+D_{\nu f}+D_{\nu\mu}+D_{\nu\nu}
\end{eqnarray}
We use $D_f$ term as an
example. From Eq.\ (\ref{eq:51}), after a natural selection process, one sequence $S_b$ is
replaced by sequence $S_a$. If it is replaced by itself, nothing is changed. Otherwise, as the initial sequences is totally random, the number of
sites changed is on average $L/2$, and the probability of this is $\langle f(a)\rangle
(1-1/N)=2L(1-1/N)$. As in the whole population, only this sequence is
changed, the divergence can be calculated as
\begin{eqnarray}
D_f &=& \frac{1}{N}\left<
\frac{L - S_b(t) \cdot S_b(0)}{2}\right>\nonumber
\\&=&2L(1-1/N)\frac{L-(L/2-L/2)}{2}\nonumber
\\&=&L^2(1-1/N)
\end{eqnarray}

Similarly, we can obtain other terms,
\begin{eqnarray}
D_{\mu}&=&\mu L\nonumber
\\D_{\nu}&=&\nu L(1-1/N)/2\nonumber
\\D_{ff}&=&-2L^3(1-1/N)+L^2(1-1/N)(1-3/N)\nonumber
\\D_{f\mu}&=&D_{\mu f}=-2\mu L^2(1-1/N)\nonumber
\\D_{f\nu}&=&D_{\nu f}=-\nu L^2(1-1/N)\nonumber
\\D_{\mu\mu}&=&-2\mu^2L\nonumber
\\D_{\mu\nu}&=&D_{\nu\mu}=-\mu\nu L(1-1/N)\nonumber
\\D_{\nu\nu}&=&-\nu^2L(1-1/N)/2
\end{eqnarray}
Adding these terms together gives Eq.\ (\ref{4}).
\clearpage

\bibliographystyle{unsrt}
\bibliography{f}

\begin{thebibliography}{10}

\bibitem{Waddington}
C.~H. Waddington.
\newblock Canalization of development and the inheritance of acquired
  characters.
\newblock {\em Nature}, 150:563--565, 1942.

\bibitem{Simon}
H.~A. Simon.
\newblock The architecture of complexity.
\newblock {\em Proc. Amer. Phil. Soc.}, 106:467--482, 1962.

\bibitem{Hartwell1999}
L.~H. Hartwell, J.~J. Hopfield, S.~Leibler, and A.~W. Murray.
\newblock From molecular to modular cell biology.
\newblock {\em Nature}, 402:C47--C52, 1999.

\bibitem{Rojas}
R.~Rojas.
\newblock {\em Neural Networks: A Systematic Introduction}.
\newblock Springer, New York, 1996.
\newblock Ch 16.

\bibitem{hgt1}
E.~Grohmann.
\newblock Horizontal gene transfer between bacteria under natural conditions.
\newblock In I~Ahmad, F.~Ahmad, and J.~Pichtel, editors, {\em Microbes and
  Microbial Technology}, pages 163--187. Springer New York, 2011.

\bibitem{hgt}
M.~B. Gogarten, J.~P. Gogarten, and L.~Olendzenski, editors.
\newblock {\em Horizontal Gene Transfer: Genomes in Flux}.
\newblock Methods in Molecular Biology, Vol. 532. Humana Press, New York, 2009.

\bibitem{Goldenfeld2011a}
N.~Chia and N.~Goldenfeld.
\newblock Statistical mechanics of horizontal gene transfer in evolutionary
  ecology.
\newblock {\em J. Stat. Phys.}, 142:1287--1301, 2011.

\bibitem{Breen2012}
M.~S. Breen et~al.
\newblock Epistasis as the primary factor in molecular evolution.
\newblock {\em Nature}, 490:535--538, 2012.

\bibitem{Sun}
J.~Sun and M.~W. Deem.
\newblock Spontaneous emergence of modularity in a model of evolving
  individuals.
\newblock {\em Phys. Rev. Lett.}, 99:228107, 2007.

\bibitem{Alon}
N.~Kashtan, M.~Parter, E.~Dekel, A.~E. Mayo, and U.~Alon.
\newblock Extinctions in heterogeneous environments and the evolution of
  modularity.
\newblock {\em Evolution}, 63:1964--1975, 2009.

\bibitem{Park06}
J.-M. Park and M.~W. Deem.
\newblock Schwinger boson formulation and solution of the {C}row-{K}imura and
  {E}igen models of quasispecies theory.
\newblock {\em J. Stat. Phys.}, 125:975--1015, 2006.

\bibitem{Levine}
E.~Cohen, D.~A. Kessler, and H.~Levine.
\newblock Recombination dramatically speeds up evolution of finite populations.
\newblock {\em Phys. Rev. Lett.}, 94:098102, 2005.

\bibitem{Shraiman}
S.~Goyal, D.~J. Balick, E.~R. Jerison, R.~A. Neher, B.~I. Shraiman, and M.~M.
  Desai.
\newblock Dynamic mutation–selection balance as an evolutionary attractor.
\newblock {\em Genetics}, 191:1309--1319, 2012.

\bibitem{Derrida1991}
B.~Derrida and L.~Peliti.
\newblock Evolution in a flat fitness landscape.
\newblock {\em Bull. Math. Biol.}, 53:355--382, 1991.

\bibitem{Peliti2011}
G.~Bianconi, D.~Fichera, S.~Franz, and L.~Peliti.
\newblock Modeling microevolution in a changing environment: the evolving
  quasispecies and the diluted champion process.
\newblock {\em J. Stat. Mech.}, 2011:P08022, 2011.

\bibitem{Park}
J.-M. Park and M.~W. Deem.
\newblock Phase diagrams of quasispecies theory with recombination and
  horizontal gene transfer.
\newblock {\em Phys. Rev. Lett.}, 98:058101, 2007.

\bibitem{Munoz2}
E.~T. Munoz, J.-M. Park, and M.~W. Deem.
\newblock Quasispecies theory for horizontal gene transfer and recombination.
\newblock {\em Phys. Rev. E}, 78:061921, 2008.

\bibitem{DS1}
Z.~Avetisyan and D.~B. Saakian.
\newblock Recombination in one- and two-dimensional fitness landscapes.
\newblock {\em Phys. Rev. E}, 81:051916, 2010.

\bibitem{DS2}
D.~B. Saakian.
\newblock Evolutionary advantage via common action of recombination and
  neutrality.
\newblock {\em Phys. Rev. E}, 88:052717, 2013.

\bibitem{DS3}
D.~B. Saakian, Z.~Kirakosyan, and C.~K. Hu.
\newblock Biological evolution in a multidimensional fitness landscape.
\newblock {\em Phys. Rev. E}, 86:031920, 2012.

\bibitem{Dotsenko}
V.~S. Dotsenko.
\newblock Fractal dynamics of spin glasses.
\newblock {\em J. Phys. C: Solid State Phys.}, 18:6023--6031, 1985.

\bibitem{Park13}
J.-M. Park, L.~R. Niestemski, and M.~W. Deem.
\newblock Quasispecies theory for evolution of modularity.
\newblock page DOI: 10.1103/PhysRevE.00.002700, 2014.
\newblock arXiv:1211.5646.

\bibitem{swine}
G.~J.~D. Smith, D.~Vijaykrishna, J.~Bahl, S.~J. Lycett, M.~Worobey, O.~G.
  Pybus, S.~K. Ma, C.~L. Cheung, J.~Raghwani, S.~Bhatt, J.~S.~Malik Peiris,
  Y.~Guan, and A.~Rambaut.
\newblock Origins and evolutionary genomics of the 2009 swine-origin h1n1
  influenza a epidemic.
\newblock {\em Nature}, 459:1122--1125, 2009.

\bibitem{Holmes}
A.~Rambaut, O.~G. Pybus, M.~I. Nelson, C.~Viboud, J.~K. Taubenberger, and E.~C.
  Holmes.
\newblock The genomic and epidemiological dynamics of human influenza a virus.
\newblock {\em Nature}, 453:615--619, 2008.

\bibitem{Marshall}
N.~Marshall, L.~Priyamvada, Z.~Ende, J.~Steel, and A.~C. Lowen.
\newblock Influenza virus reassortment occurs with high frequency in the
  absence of segment mismatch.
\newblock {\em PLoS Pathog.}, 9:e1003421, 2013.

\bibitem{Trifonov}
V.~Trifonov, H.~Khiabanian, and Raul Rabadan.
\newblock Geographic dependence, surveillance, and origins of the 2009
  influenza a (h1n1) virus.
\newblock {\em N. Engl. J. Med.}, 361:115--119, 2009.

\bibitem{Lassig}
M.~{\L}uksza and M.~L{\"a}ssig.
\newblock A predictive fitness model for influenza.
\newblock {\em Nature}, 507:57--61, 2014.

\bibitem{Smith2}
D.~J. Smith, A.~S. Lapedes, J.~C. de~Jong, T.~M. Bestebroer, G.~F. Rimmelzwaan,
  A.~D. M.~E. Osterhaus, and R.~A.~M. Fouchier.
\newblock Mapping the antigenic and genetic evolution of influenza virus.
\newblock {\em Science}, 305:371--376, 2004.

\bibitem{Gupta}
V.~Gupta, D.~J. Earl, and M.~W. Deem.
\newblock Quantifying influenza vaccine efficacy and antigenic distance.
\newblock {\em Vaccine}, 24:3881--3888, 2006.

\bibitem{He}
J.~He and M.~W. Deem.
\newblock Low-dimensional clustering detects incipient dominant influenza
  strain clusters.
\newblock {\em PEDS}, 23:935--946, 2010.

\bibitem{deem2003sequence}
M.~W. Deem and H.-Y. Lee.
\newblock Sequence space localization in the immune system response to
  vaccination and disease.
\newblock {\em Phys. Rev. Lett.}, 91:068101, 2003.

\bibitem{wu1982potts}
F.~Y. Wu.
\newblock The potts model.
\newblock {\em Rev. Mod. Phys.}, 54:235--268, 1982.

\bibitem{Munoz}
E.~T. Munoz, J.-M. Park, and M.~W. Deem.
\newblock Solution of the crow-kimura and eigen models for alphabets of
  arbitrary size by schwinger spin coherent states.
\newblock {\em J. Stat. Phys}, 135:429--465, 2009.

\bibitem{wigner1958distribution}
P.~E. Wigner.
\newblock On the distribution of the roots of certain symmetric matrices.
\newblock {\em Annals of Mathematics}, pages 325--327, 1958.

\bibitem{li2001gaussian}
W.~V. Li and Q.-M. Shao.
\newblock Gaussian processes: inequalities, small ball probabilities and
  applications.
\newblock {\em Stochastic processes: theory and methods}, 19:533--597, 2001.

\bibitem{PhysRevB.76.184412}
S.-Y. Kim, S.-J. Lee, and J.~Lee.
\newblock Ground-state energy and energy landscape of the
  sherrington-kirkpatrick spin glass.
\newblock {\em Phys. Rev. B}, 76:184412, 2007.

\bibitem{0305-4470-30-24-010}
V.~M. de~Oliveira and J.~F. Fontanari.
\newblock Landscape statistics of the $p$-spin ising model.
\newblock {\em Journal of Physics A: Mathematical and General}, 30:8445--8457,
  1997.

\bibitem{grest1983ground}
G.~S. Grest and C.~M. Soukoulis.
\newblock Ground-state properties of the infinite-range vector spin-glasses.
\newblock {\em Physical Review B}, 28:2886--2889, 1983.

\bibitem{PhysRevLett.36.1217}
J.~M. Kosterlitz, D.~J. Thouless, and R.~C. Jones.
\newblock Spherical model of a spin-glass.
\newblock {\em Phys. Rev. Lett.}, 36:1217--1220, 1976.

\bibitem{pisier1999volume}
G.~Pisier.
\newblock {\em The volume of convex bodies and Banach space geometry},
  volume~94.
\newblock Cambridge University Press, New York, 1999.

\bibitem{owen1962moments}
D.~B. Owen and G.~P. Steck.
\newblock Moments of order statistics from the equicorrelated multivariate
  normal distribution.
\newblock {\em The Annals of Mathematical Statistics}, 33:1286--1291, 1962.

\end{thebibliography}

\end{document}